\documentclass[
 reprint,
 amsmath,amssymb,
 aps,
pra,
]{revtex4-2}
\usepackage{graphicx}
\usepackage{float}
\usepackage{dcolumn}
\usepackage{bm}
\usepackage{bbm}
\usepackage{braket}
\usepackage{xcolor}
\usepackage[T1]{fontenc}
\usepackage{enumerate,enumitem}
\usepackage{booktabs}
\usepackage{physics}
\usepackage{amsmath}
\usepackage{appendix}
\usepackage{hyperref}

\makeatletter
\DeclareFontFamily{OMX}{MnSymbolE}{}
\DeclareSymbolFont{MnLargeSymbols}{OMX}{MnSymbolE}{m}{n}
\SetSymbolFont{MnLargeSymbols}{bold}{OMX}{MnSymbolE}{b}{n}
\DeclareFontShape{OMX}{MnSymbolE}{m}{n}{
    <-6>  MnSymbolE5
   <6-7>  MnSymbolE6
   <7-8>  MnSymbolE7
   <8-9>  MnSymbolE8
   <9-10> MnSymbolE9
  <10-12> MnSymbolE10
  <12->   MnSymbolE12
}{}
\DeclareFontShape{OMX}{MnSymbolE}{b}{n}{
    <-6>  MnSymbolE-Bold5
   <6-7>  MnSymbolE-Bold6
   <7-8>  MnSymbolE-Bold7
   <8-9>  MnSymbolE-Bold8
   <9-10> MnSymbolE-Bold9
  <10-12> MnSymbolE-Bold10
  <12->   MnSymbolE-Bold12
}{}

\let\llangle\@undefined
\let\rrangle\@undefined
\DeclareMathDelimiter{\llangle}{\mathopen}%
                     {MnLargeSymbols}{'164}{MnLargeSymbols}{'164}
\DeclareMathDelimiter{\rrangle}{\mathclose}%
                     {MnLargeSymbols}{'171}{MnLargeSymbols}{'171}
\makeatother

\begin{document}

\title{Loss-tolerant parallelized Bell-state generation with a hybrid cat qudit}

\author{Z.~M.~McIntyre$^{1,2}$}
\email{zoe.mcintyre@unibas.ch}
\author{W.~A.~Coish$^{2}$}%
 \email{william.coish@mcgill.ca}
\affiliation{$^{1}$Department of Physics, University of Basel, Klingelbergstrasse 82, 4056 Basel, Switzerland}
\affiliation{%
 $^{2}$Department of Physics, McGill University, 3600 rue University, Montreal, QC, H3A 2T8, Canada
}%

\date{\today}

\begin{abstract}
    Having multiple Bell pairs shared by distant quantum registers provides a key resource for both quantum networks and distributed quantum computing. In this paper, we present a protocol for parallelized Bell-pair generation that uses the phase of a coherent light pulse to encode a qudit, enabling the simultaneous generation of multiple Bell pairs. By encoding a qudit in a basis of light-matter Schrödinger's cat states, the loss of a photon in transit can be detected through an $XX$ parity syndrome, allowing the backaction due to the lost photon to be deterministically corrected through single-qubit rotations. The protocol presented here is compatible with existing technologies in both optical and microwave (circuit QED) architectures, supporting near-term implementation across diverse quantum platforms.
\end{abstract}

\maketitle

\section{Introduction}

The ability to entangle distant qubits is a useful primitive for both quantum networks and distributed quantum computing~\cite{azuma2023quantum}.
Such entanglement could be established by applying quantum logic gates designed to operate over longer distances~\cite{duan2005robust,daiss2021quantum,mcintyre2025protocols}, or by using strategies tailored for the creation of specific entangled resource states such as Bell states~\cite{cirac1997quantum,duan2001long,barrett2005efficient}. In an architecture consisting of groups of stationary qubits (nodes) connected by photonic interconnects (channels), the clock speed of inter-node operations could be improved by making efficient use of the channels---for instance, by generating multiple Bell pairs with a single channel use. The creation of multiple Bell pairs is a prerequisite for entanglement purification protocols~\cite{bennett1996purification,deutsch1996quantum,pan2001entanglement}, which could be used to realize high-fidelity quantum communication over potentially noisy channels.  In combination with gate teleportation~\cite{gottesman1999demonstrating}, Bell states can also be used to apply entangling gates between spatially separated qubits, as has been demonstrated experimentally~\cite{chou2018deterministic,wan2019quantum,feng2025chip,main2025distributed}. In the context of quantum error correction, generating several Bell pairs with a single channel use would then allow several entangling gates to be teleported in parallel. This could be useful for implementing transversal controlled-NOT gates between logical qubits encoded in spatially separated surface-code patches. 

\begin{figure}
    \centering
    \includegraphics[width=0.9\linewidth]{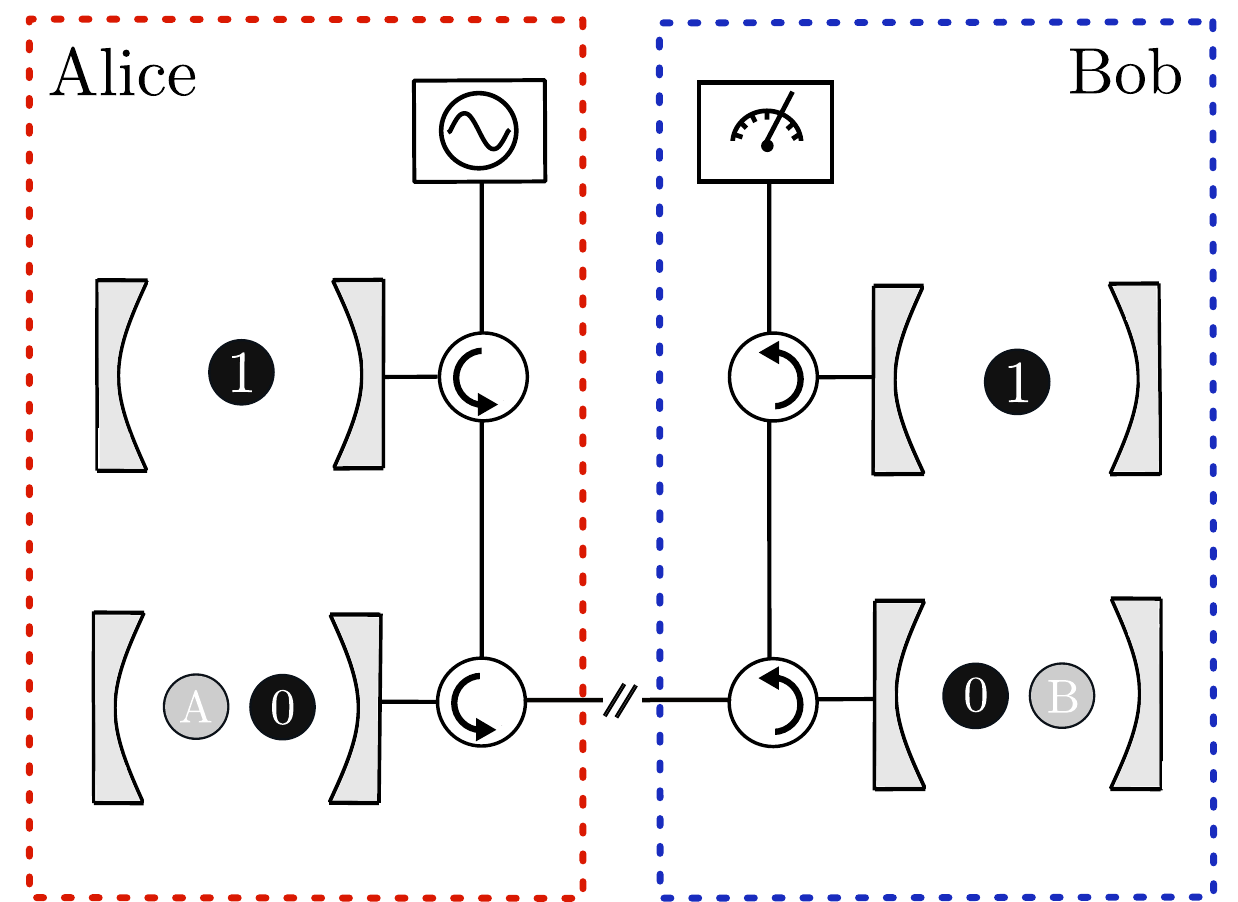}
    \caption{Schematic of the setup: Alice and Bob each possess a register of $N$ qubits numbered $0,\dots,N-1$ (here, $N=2$), with each qubit coupled to a single quantized cavity mode. In the loss-tolerant version of the protocol, Alice and Bob need one additional ancilla each, here labeled A and B, respectively (in gray). A coherent light pulse prepared by Alice becomes entangled first with Alice's qubits then with Bob's qubits through a series of qubit-state-conditioned phase shifts imparted upon reflection from each cavity in turn. A heterodyne measurement of the light pulse, together with a circuit applied locally on Bob's register, can then be used to produce $N$ Bell pairs shared by Alice and Bob. The loss of a photon in transit can be detected by also entangling the light pulse with the ancillas A and B. In this case, a measurement of the ancilla-qubit $XX$ parity will reveal whether a photon was lost. Conditioned on an eigenvalue $XX=-1$, Alice can correct the backaction due to the loss by applying single-qubit rotations to her register. The ancillas could either be coupled to their own separate cavities or to the same cavities as one of Alice's and Bob's register qubits (as pictured here).}
    \label{fig:setup}
\end{figure}

Two or more Bell pairs can be generated with the exchange of a single photon by encoding more than one qubit in separate photonic degrees of freedom---a strategy known as quantum multiplexing~\cite{lo2019quantum,zhou2023parallel,xie2023heralded}.  Rather than encode one or more qubits, the Hilbert space of a single photon can also be used to encode a \textit{qudit}, and the simultaneous generation of $N$ Bell pairs through the exchange of a time-bin qudit has been proposed as well~\cite{xie2021quantum,zheng2022entanglement}. These strategies, being based on the transmission of single photons, are non-deterministic in the presence of photon loss, but their successful implementation can be heralded by the detection of the photon at the end of the protocol. 

In this paper, we show how $N$ Bell pairs can be generated by encoding information in the  phase degree-of-freedom of a propagating coherent light pulse. The use of a multiphoton state allows errors due to the loss of a photon to be detected and corrected, potentially helping to mitigate the loss-related heralding delays commonly associated with entanglement-generation schemes based on the exchange of single photons, without any need for quantum memories as in multiplexed single-photon protocols~\cite{collins2007multiplexed,simon2007quantum,van2017multiplexed,ruskuc2025multiplexed}.  The use of bosonic cat codes~\cite{bergmann2016quantum} for detecting photon loss has been studied for protocols resulting in a single Bell state~\cite{roy2016concurrent,li2023memoryless}, and similar ideas underpin bosonic qubit encodings based on superpositions of cat states~\cite{mirrahimi2014dynamically}. Here, we show how parallelized Bell-state generation can be made tolerant to the loss of a photon by encoding a qudit in a basis of light-matter Schrödinger's cat states, then sending the photonic portion of the qudit from one node--``Alice''--to a second distant node--``Bob'' (Fig.~\ref{fig:setup}). This strategy would enable multiple Bell pairs to be created using elements that are routinely realized for both optical and microwave setups: coherent light sources, cavity-qubit coupling, heterodyne detection, and high-fidelity qubit gates. For microwave-regime circuit QED implementations in particular, the scheme proposed here circumvents the technical challenges of performing operations on a time-bin degree-of-freedom, which typically requires delay lines. One potential additional advantage of leveraging a phase degree-of-freedom as part of the qudit encoding is that the total photon-pulse duration does not scale exponentially in the number $M$ of qubits, as it would for schemes mediated by time-bin qudits having $2^M$ time-bin states.

The layout of this paper is as follows: In Sec.~\ref{sec:protocol}, we lay out the steps of the protocol in the absence of errors. Two qudit encodings are considered: The first encodes the qudit in the phase of the coherent state, while the second enables loss correction by encoding the qudit in a hybrid light-matter basis. Errors are considered in more detail in Sec.~\ref{sec:errors}, where we consider imperfections due to multiphoton losses, non-orthogonality of the qudit basis states, and dephasing of stationary qubits.  The paper concludes in Sec.~\ref{sec:conclusions}.

\section{Protocol}\label{sec:protocol}

In Sec.~\ref{sec:phase-qudit}, we first give a strategy for generating multiple Bell pairs with a qudit encoded in the phase of a coherent state. With this first protocol, photon loss will contribute to the state infidelity at linear order in the average number of lost photons since loss events cannot be detected. We then show in Sec.~\ref{sec:cat-qudit} how the same operations can be used to encode the qudit in a basis of light-matter (hybrid) cat states that allow the photon-number parity of the light pulse to be entangled with the joint $XX$ parity of a pair of ancilla qubits held by Alice and Bob. A change in the photon-number parity leads to an error syndrome consisting of an eigenvalue $XX=-1$, which can be corrected through single-qubit rotations of Alice's qubits to ensure the correct post-measurement state. This section focuses on introducing the required operations in the absence of imperfections, dominant sources of which are discussed and modelled in Sec.~\ref{sec:errors}.

\subsection{Phase encoding}\label{sec:phase-qudit}

\begin{figure}
    \centering
    \includegraphics[width=0.8\linewidth]{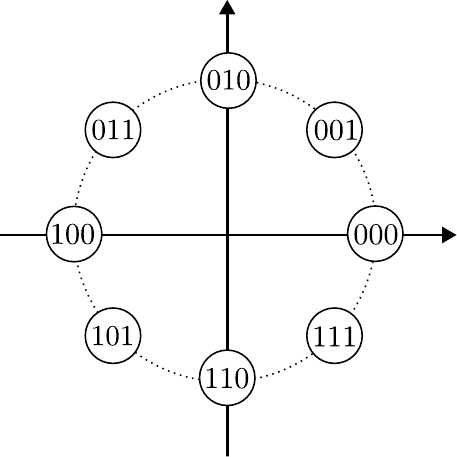}
    \caption{Phase-space representation of the entangling operation for a phase-encoded qudit and $N=3$: For each of the $2^N=8$ basis states of Alice's qubits, an initial coherent state $\ket{\alpha}$ undergoes a phase-space rotation by an amount $m\varphi$, where $\varphi=2\pi/2^{N}$ and $m=\sum_{j=0}^2 s_i 2^i$ is the base-10 representation of the computational basis state $\ket{s_2 s_1 s_0}$ ($s_i=0,1$). }
    \label{fig:entangling-op}
\end{figure}

We consider a setup where Alice and Bob each have a register of $N$ qubits as illustrated in Fig.~\ref{fig:setup}. (The ancillas A and B also depicted in Fig.~\ref{fig:setup} are not needed at present and may be disregarded). The protocol begins with all $2N$ qubits initialized in the state $\ket{+}$, where $\ket{+}=(\ket{0}+\ket{1})/\sqrt{2}$. A propagating light pulse prepared in a coherent state $\ket{\alpha}$ is first entangled with Alice's register through an entangling operation of the form
\begin{equation}\label{phase-shifts}
    \ket{m}_{\mathrm{A}}\ket{\alpha}\mapsto \ket{m}_{\mathrm{A}}\ket*{e^{im \varphi}\alpha},
\end{equation}
where here, $\ket{m}_{\mathrm{A}}=\ket{s_{N-1},\dots,s_0}_{\mathrm{A}}$ is a computational basis state of Alice's register labeled by its decimal representation $m=\sum_{j=0}^{N-1} s_j 2^j$, and where $\varphi$ is a fixed angular increment given by
\begin{equation}\label{blob-sep}
    \varphi=\frac{2\pi}{2^{N}}.
\end{equation} 
The entangling operation given in Eq.~\eqref{phase-shifts} will rotate the initial coherent state $\ket{\alpha}$ in phase space by a different amount for each state $\ket{m}$ of Alice's register (Fig.~\ref{fig:entangling-op}). The total phase shift $m\varphi$ acquired conditionally on Alice's register being in state $\ket{m}$ can be decomposed into a sum of phase shifts $\varphi_j=2^j \varphi$, each imparted conditioned on the state of Alice's $j^{\mathrm{th}}$ qubit: 
\begin{align}
\begin{aligned}\label{phase-breakdown}
    m\varphi=\sum_{j=0}^{N-1}s_j \varphi_j,\quad \varphi_j=2^j\varphi.
\end{aligned}
\end{align}
Assuming dispersive coupling with strength $\chi_j$ of qubit $j$ to its own cavity, the entangling operation of Eq.~\eqref{phase-shifts} can be realized (up to a global phase, which can be compensated through a redefinition of the zero-phase reference of the light pulse) by setting $\chi_j=(\kappa_j/2)\tan{(\varphi_j/4)}$, 
where here, $\kappa_j$ is the decay rate of the cavity mode coupled to qubit $j$. This result is derived in Appendix \ref{sec:entangling-operation}. In order for these phase shifts to be well defined, the spatial extent of the light pulse must be large relative to $v/\kappa_{\mathrm{min}}$, where $v$ is the speed of light and $\kappa_{\mathrm{min}}$ is the smallest of the $\kappa_j$. Well-defined qubit-state-conditioned phase shifts of $\pi$ based on dispersive coupling have been realized experimentally~\cite{kono2018quantum,wang2022flying}. They have also been realized using strong resonant coupling between a cavity and a pair of levels in a three-level system encoding a qubit~\cite{besse2018single, hacker2019deterministic, besse2020parity}.

Starting from the state $\ket{+}^{\otimes N}\ket{\alpha}$, the entangling operation  given in Eq.~\eqref{phase-shifts} can be used to generate entanglement between Alice's register and a phase-encoded qudit with basis states $\ket{k}_{\mathrm{phase}}=\ket*{e^{ik\varphi}\alpha}$ ($k=0,\dots,2^N-1$), corresponding to the state 
\begin{equation}\label{state-transmitted}
    \ket{\Psi}_{\mathrm{A}}=\frac{1}{\sqrt{2^N}}\sum_{m=0}^{2^N-1}\ket{m}_{\mathrm{A}}\ket*{m}_{\mathrm{phase}}.
\end{equation} 
Note that while Eq.~\eqref{state-transmitted} gives the produced state even for small amplitudes $\alpha$, this state only approaches a maximally entangled state in the limit  (considered in this paper) where the qudit basis states are near-orthogonal, requiring $\alpha$ such that $\langle k \vert k'\rangle_{\mathrm{phase}}\approx \delta_{k,k'}$. The corrections to this approximation are exponentially suppressed in $\vert \alpha\vert^2$, and their impact on measurement errors is discussed in Sec.~\ref{sec:errors}.
The light pulse encoding the phase qudit is then sent to Bob, where it acquires an additional phase shift conditioned on the state $\ket{m}_{\mathrm{B}}$ of Bob's register according to
\begin{equation}\label{entangle-Bob}
    \ket{m}_{\mathrm{B}}\ket{\alpha}\mapsto\ket{m}_{\mathrm{B}}\ket*{e^{-im\varphi}\alpha}.
\end{equation}
Relative to Eq.~\eqref{phase-shifts}, the only difference is that for Bob, the $m$-conditioned phase shift differs by a sign and is given by $-m\varphi$ rather than $m\varphi$. Following the interaction of the light pulse with Bob's register, the final state $\ket{\Psi}$ of Alice's register, Bob's register, and the phase qudit is given by
\begin{equation}\label{pre-meas-state}
    \ket{\Psi}=\frac{1}{2^N}\sum_{m,n=0}^{2^N-1}\ket{m}_{\mathrm{A}}\ket{n}_{\mathrm{B}}\ket*{m-n}_{\mathrm{phase}}.
\end{equation}

In order to disentangle the phase qudit in a way that leaves behind maximal entanglement between Alice and Bob's registers, the qudit must be measured in the computational basis $\ket{m}_{\mathrm{phase}}$. Such a projection onto the computational basis can be accomplished straightforwardly using heterodyne detection.
This measurement will generally lead to errors related to the non-orthogonality of the qudit basis states for finite $\alpha$, but we treat it as ideal for now by approximating $\langle k\vert k'\rangle_{\mathrm{phase}}\approx \delta_{k,k'}$. Under this approximation, and for the measurement outcome $\ket{-k}_{\mathrm{phase}}$, the post-measurement state $\ket{\Psi_k}$ of Alice and Bob's registers is given by
\begin{equation}\label{ideal-almost-bell}
    \ket{\Psi_k}=\frac{1}{\sqrt{2^N}}\sum_{m=0}^{2^N-1}\ket{m}_{\mathrm{A}}\ket{m \oplus k}_{\mathrm{B}},
\end{equation}
where $\oplus$ denotes addition modulo $2^N$. For $k=0$, $\ket{\Psi_0}$ is a product of Bell states $\ket{\Phi^+}\propto \ket{00}+\ket{11}$ shared by Alice and Bob, $\ket{\Psi_0}=\ket{\Phi^+}^{\otimes N}$, while for $k=2^{N-1}$, $\ket{\Psi_{2^{N-1}}}$ can be transformed into $\ket{\Psi_0}$ by applying a bitflip to Bob's qubit $j=N-1$. For $k\neq 0,2^{N-1}$, $\ket{\Psi_k}$ cannot generally be factorized into a product of Bell states shared by Alice and Bob. However, given knowledge of the measurement outcome $\ket{-k}_{\mathrm{phase}}$, Bob can perform operations on his register to transform $\ket{\Psi_k}$ into $\ket{\Psi_0}$. The practical significance of this statement is that once $\ket{\Psi_k}$ has been obtained, no non-local operations between Alice and Bob are required to obtain $N$ copies of the Bell state $\ket{\Phi^+}$. 

Formally, the operation to be implemented by Bob is a Boolean operation taking $\ket{m}\mapsto \ket{m\oplus (-k)}$. A procedure for realizing this operation is as follows~\cite{draper2000addition}: First, Bob applies a quantum Fourier transform to his qubits via some unitary operator $U_{\mathrm{QFT}}$. Acting on state $\ket{m}$, $U_{\mathrm{QFT}}$ produces the state 
\begin{equation}\label{qft}
    U_{\mathrm{QFT}}\ket{m}=\frac{1}{\sqrt{2^N}}\sum_{n=0}^{2^N-1} \omega_N^{mn}\ket{n},
\end{equation}
where here, $\omega_N=e^{\frac{2\pi i}{2^N}}$ is the $2^N$-th root of unity. Next, Bob applies to his qubits the unitary operation $R^k$, where
\begin{equation}\label{r-operator}
    R=\prod_j R_j(\phi_j).
\end{equation}
Here, $R_j(\phi_{j})=e^{-i\phi_{j}}\ketbra{1}_j$ is a single-qubit phase gate with $\phi_j=2\pi/2^j$ for $j=0,\dots,N-1$~\cite{draper2000addition}. This phase gate could be applied in software through a change of phase reference~\cite{mckay2017efficient}. Applied to state $\ket{n}$ in Eq.~\eqref{qft}, the unitary $R^k=\prod_{j}R_j^k(\phi_{j})$ acts according to $R^k\ket{n}=\omega_N^{-kn}\ket{n}$. The last step is for Bob to invert the quantum Fourier transform through the application of $U_{\mathrm{QFT}}^\dagger$, thereby completing the mapping $\ket{m}\mapsto\ket{m\oplus (-k)}$. Although circuits implementing the quantum Fourier transform typically assume all-to-all connectivity, $U_{\mathrm{QFT}}$ could also be implemented in a nearest-neighbour architecture using the circuit given in Ref.~\cite{fowler2004implementation}. 


\subsection{Cat encoding}\label{sec:cat-qudit}

In Sec.~\ref{sec:phase-qudit}, we focused on describing the protocol in the absence of photon loss (and other implementation errors). However, the light pulse encoding the qudit may lose photons while traveling from Alice to Bob, resulting in a state infidelity that scales linearly (for $n_\ell<1$) in the average number $n_\ell$ of lost photons. This infidelity can be pushed out to $O(n_\ell^2)$ by using a different qudit encoding that instead allows errors due to the loss of a photon to be detected and  corrected: Rather than encode a qudit in the phase of the coherent state, the qudit can alternatively be encoded in a basis $\{\ket{k}_{\mathrm{cat}}\}_{k=0}^{2^N-1}$ of light-matter (hybrid) cat states given by 
\begin{equation}\label{cat-qudit-basis}
    \ket{k}_{\mathrm{cat}}=\frac{1}{\sqrt{2}}\left(\ket*{\alpha e^{ik\varphi}}\ket{0}+\ket*{-\alpha e^{ik\varphi}}\ket{1}\right),
\end{equation}
where here, $\ket{0}$ and $\ket{1}$ are the computational basis states of an ancilla qubit held by Alice. Since $e^{i(k+2^{N-1})\varphi}=-e^{ik\varphi}$, it is worth remarking that $k$ and $k+2^{N-1}$ do indeed label distinct states: The orthogonality of the ancilla-qubit basis states is what ensures that $\ket{k}_{\mathrm{cat}}$ and $\ket*{k+2^{N-1}}_{\mathrm{cat}}$ are approximately orthogonal for large $\alpha$, up to corrections that are exponentially small in $2\lvert\alpha\rvert^2$. 

For this protocol, Bob holds an ancilla qubit as well (Fig.~\ref{fig:setup}), but apart from the two ancilla qubits held by Alice and Bob, the hardware requirements for the cat-qudit-mediated protocol are the same as for the phase-qudit-mediated protocol: The light pulse is initially prepared in a coherent state and measured with heterodyne detection.

\subsubsection{Ideal scenario}

As before, we begin by laying out the protocol in the absence of errors. A light pulse prepared in a coherent state $\ket{\alpha}$ first becomes entangled with Alice's qubits by acquiring a qubit-state-conditioned phase shift according to 
\begin{equation}\label{alice-entangle-cat}
    \ket{m,s}_{\mathrm{A}}\ket{\alpha}\mapsto\ket{m,s}_{\mathrm{A}}\ket*{(-1)^{s}e^{im\varphi}\alpha},
\end{equation}
where here, $s=0,1$ labels the computational basis state $\ket{s}$ of Alice's ancilla qubit. This ancilla could be coupled to its own cavity, or---to reduce the total number of cavities---to the same cavity as the last of Alice's $N$ qubits to interact with the light pulse before it is sent to Bob, which we take to be qubit $j=0$ for concreteness (Fig.~\ref{fig:setup}). 
In the latter scenario, the light pulse would acquire a phase shift conditioned on the joint state of \textit{both} Alice's ancilla qubit and qubit $j=0$. With dispersive coupling to a common cavity mode, the two-qubit-conditioned phase shift required to realize Eq.~\eqref{alice-entangle-cat} can be generated, up to a redefinition of the phase reference, by tuning the dispersive shifts $\chi_{\mathrm{A}}$ and $\chi_{0}$ of the ancilla and qubit $j=0$ so that $\chi_{\mathrm{A}}=-\kappa_{0}\sec{(\varphi_0/2)}$ and $\chi_{0}=\kappa_{0}\tan{(\varphi_0/2)}$ (Appendix \ref{sec:entangling-operation}).  

With all of Alice's qubits (including the ancilla) prepared in $\ket{+}$, the joint state $\ket{\Phi}_{\mathrm{A}}$ of the qubits and light pulse obtained under the entangling operation of Eq.~\eqref{alice-entangle-cat} is given by 
\begin{align}\label{send-parity}
\begin{aligned}
    \ket{\Phi}_{\mathrm{A}}&=\frac{1}{\sqrt{2^{N}}}\sum_{m=0}^{2^N-1}\ket{m}\ket{m}_{\mathrm{cat}}\\
    &=\frac{1}{\sqrt{2^N}}\sum_{m}\sum_{\sigma=\pm}\frac{\mathcal{N}_\alpha^\sigma}{2}\ket{m}\ket*{C_{\alpha e^{im\varphi}}^\sigma}\ket{\sigma},
\end{aligned}
\end{align}
where in the second line, we have introduced the cat states $\ket{C_\alpha^\pm}=(\mathcal{N}_\alpha^\pm)^{-1}(\ket{\alpha}\pm\ket{-\alpha})$ with $\mathcal{N}_\alpha^\pm=[2(1\pm e^{-2\lvert\alpha\rvert^2})]^{1/2}$, and where $\ket{\sigma=\pm}$ are Pauli-X eigenstates. From the first equality above, we see that Alice's qubits become maximally entangled with the hybrid cat qudit whose basis states were given in Eq.~\eqref{cat-qudit-basis}. Since the (photonic) cat states $\ket*{C_{\alpha e^{im\varphi}}^\pm}$ are states of definite photon-number parity, the second equality in Eq.~\eqref{send-parity} makes it clear that with this choice of qudit encoding, the photon-number parity of the light pulse sent to Bob is correlated with the $X$-basis eigenstate $\ket{\sigma=\pm}$ of Alice's ancilla qubit. This correlation is what will ultimately enable the errors caused by a lost photon to be detected and corrected. Since both states of the form $\ket{k}_{\mathrm{cat}}$ [Eq.~\eqref{cat-qudit-basis}], involving maximal entanglement between two systems, and equal-superposition states of the form $\ket{C_\alpha^\pm}$ are referred to as ``cat states'' in the literature, we will attempt to resolve any ambiguity by referring to them as hybrid and photonic cat states, respectively, in cases where the intended meaning is not immediately clear from context.  

The light pulse is sent to Bob, and as before, the phase shift $-m\varphi$ acquired conditionally on Bob's register being in state $\ket{m}_{\mathrm{B}}$ is equal in magnitude but opposite in sign to the phase shift acquired conditionally on Alice's qubits being in the same state $\ket{m}_{\mathrm{A}}$. The main difference relative to the phase-qudit-mediated protocol [cf.~Eq.~\eqref{entangle-Bob}] is that upon arriving at the location of Bob's register, the light pulse will also acquire a $\pi$ phase shift conditioned on Bob's ancilla qubit being in state $\ket{1}$, so that the full entangling operation for Bob's qubits reads
\begin{equation}\label{entangle-bob-2}
\ket{m,s}_{\mathrm{B}}\ket{\alpha}\mapsto \ket{m,s}_{\mathrm{B}}\ket{(-1)^s e^{-im\varphi}\alpha}.    
\end{equation}
Under this map, Bob's ancilla qubit (prepared in $\ket{+}$) will undergo a phase flip conditioned on the photon-number parity of the photonic cat state, which will itself undergo an $m$-dependent rotation in phase space according to
\begin{align}\label{bob-cat-flip}
    \ket{m}_{\mathrm{B}}\ket{+}\ket{C_\alpha^\pm}\mapsto \ket{m}_{\mathrm{B}}\ket{\pm}\ket*{C_{\alpha e^{-im\varphi}}^\pm}.
\end{align}
In the absence of photon loss, the final state $\ket{\Phi}$ of the light pulse and all stationary qubits is therefore given by
\begin{align}
   \ket{\Phi}&=\frac{1}{2^N}\sum_{m,n=0}^{2^N-1}\ket{m}_{\mathrm{A}}\ket{n}_{\mathrm{B}}\nonumber\\
   &\times \frac{1}{\sqrt{2}}\left(\ket{m-n}_{\mathrm{cat}}\ket{0}+\ket{m-n+2^N}_{\mathrm{cat}}\ket{1}\right)\label{final-cat-1}\\
   &=\frac{1}{2^N}\sum_{\substack{m,n\\\sigma=\pm}}\ket{m}_{\mathrm{A}}\ket{n}_{\mathrm{B}}\frac{\mathcal{N}_{\alpha}^\sigma}{2}\ket*{C_{\alpha e^{i(m-n)\varphi}}^\sigma}\ket{\sigma\sigma}.\label{final-cat-2}
\end{align}
In Eq.~\eqref{final-cat-1}, $\ket{0}$ and $\ket{1}$ are the basis states of Bob's ancilla qubit (recall that Alice's ancilla qubit is also used to encode the hybrid cat qudit), while in Eq.~\eqref{final-cat-2}, $\ket{\sigma\sigma}$ is a joint state of both ancilla qubits. 

Heterodyne detection of the light pulse can again be used to implement a projection onto the basis $\{\ket{k}_{\mathrm{phase}}\}$. Note that since the state $\ket{m}_{\mathrm{A}}\ket{n}_{\mathrm{B}}$ of Alice and Bob's qubits is correlated with the phase $e^{i(m-n)\varphi}$ of a cat state [Eq.~\eqref{final-cat-2}], there are now two possible values of the phase-space rotation, i.e.~$\ket{k}_{\mathrm{phase}}$ outcome, associated with each value of $m-n$: one equal to $(m-n)\varphi$ [associated with the state $\ket{e^{i(m-n)\varphi}\alpha}$], and one equal to $(m-n-2^N/2)\varphi$ [associated with $\ket{-e^{i(m-n)\varphi}\alpha}$]. For the measurement outcome $\ket{-k}_{\mathrm{phase}}$, and neglecting measurement errors, the post-measurement state $\ket{\Phi_k}$ of the $2N+2$ stationary qubits is then given by
\begin{equation}\label{ideal-alice-bob-ancilla}
    \ket{\Phi_k}=\frac{1}{\sqrt{2}}\left(\ket{\Psi_k}\ket{\Phi_X^+}+\ket{\Psi_{k+2^{N-1}}}\ket{\Phi_X^-}\right),
\end{equation}
where $\ket{\Phi_X^\pm}=(\ket{++}\pm\ket{--})/\sqrt{2}$ are Bell states of the ancilla qubits. 

The last step of the protocol involves two measurements of the ancilla qubits: an $XX$ parity check followed by independent $Z$-basis measurements. The parity check could be realized using the strategy given in Ref.~\cite{mcintyre2024flying} for performing long-range parity checks using coherent light pulses and homodyne detection. (If the ancillas are coupled to the same cavity as a register qubit, then these register qubits must be detuned during the check.) The purpose of this parity measurement is to detect whether a photon was lost while traveling from Alice to Bob. It should be noted and acknowledged that the photon-loss probability needed to make $N$ Bell pairs must satisfy a requirement, given in Sec.~\ref{sec:errors}, that becomes increasingly stringent as $N$ is increased. This ultimately follows from the need to distinguish between $2^N$ coherent states, which must be sufficiently separated in phase space for the light pulse to encode a qudit with near-orthogonal basis states. Since the measurement of the light pulse involved in the parity check only needs to distinguish between two states, rather than $2^N$ states, we treat potential measurement errors in the parity check as negligible relative to measurement errors in the heterodyne detection of the qudit itself, which we analyze in detail in Sec.~\ref{sec:errors}. The parity check can also be implemented with a low-amplitude coherent pulse (containing on average two photons for measurement errors of 0.2\%, or three photons for measurement errors 0.02\%, for instance~\cite{mcintyre2024flying}), making errors due to photon loss during the parity check also negligible relative to photon loss during the transmission of the light pulse encoding the qudit.

In the ideal case where no photons were lost during transmission of the qudit, the parity check returns an eigenvalue of $+1$ and acts trivially on $\ket{\Phi_k}$ [Eq.~\eqref{ideal-alice-bob-ancilla}] since $\ket{\Phi_X^\pm}$ both have an even $X$-basis parity. However, in the event that a photon was lost, the parity check will return an eigenvalue of $-1$, in which case Alice should apply a correction operator (consisting of single-qubit rotations as explained in Sec.~\ref{sec:one-loss}, below) to her qubits. This correction operator effectively reverses the backaction associated with the loss of the photon. The second measurement is a measurement of both ancilla qubits in the computational ($Z$) basis. This measurement will return $\ket{\Psi_{k'}}$ [Eq.~\eqref{ideal-almost-bell}] with $k'=k$ for the measurement outcomes $\ket{00}$ and $\ket{11}$, and $k'=k+2^{N-1}$ for $\ket{01}$ and $\ket{10}$. This follows straightforwardly from the fact that $\ket{\Phi_X^+}\propto \ket{00}+\ket{11}$ while $\ket{\Phi_X^-}\propto \ket{01}+\ket{10}$.

\subsubsection{Correcting for single-photon loss}\label{sec:one-loss}

To determine the correction operator that must be applied conditioned on an $XX=-1$ outcome for the parity check, we consider a simplified loss model where the loss of a photon is modeled via the action of an annihilation operator $\hat{a}$ satisfying $\hat{a}\ket{\alpha}=\alpha\ket{\alpha}$, and where we neglect the possibility of multiphoton loss events (which are analyzed in Sec.~\ref{sec:errors}). Prior to the light pulse interacting with Bob's qubits, the state of Alice's qubits and the cat qudit is given by $\ket{\Phi}_{\mathrm{A}}$ [cf.~Eq.~\eqref{send-parity}]. Given a state with a well-defined (definite) photon-number parity, the loss of a photon will necessarily lead to a flip in the number parity. In this case, the new state $\ket{\Phi'}_{\mathrm{A}}=\hat{a}\ket{\Phi}_{\mathrm{A}}/\vert \hat{a}\ket{\Phi}_{\mathrm{A}}\rvert$ is given by
\begin{equation}\label{simplified-loss}
    \ket{\Phi'}_{\mathrm{A}}=\frac{1}{\sqrt{2^N}}\sum_{m}\sum_{\sigma=\pm}\frac{\mathcal{N}_\alpha^\sigma}{2}e^{im\varphi}\ket{m}\ket*{C_{\alpha e^{im\varphi}}^\sigma}\ket{\bar{\sigma}},
\end{equation}
where here, $\bar{\sigma}=-\sigma$. Relative to $\ket{\Phi}_{\mathrm{A}}$, the state $\ket{\Phi'}_{\mathrm{A}}$ differs in two ways: First, there is the aforementioned flip in the parity of the cat states $\ket{C_\alpha^\pm}$. Second, there is an additional $m$-dependent phase factor appearing in front of each basis state $\ket{m}$ in the superposition. For the purpose of relating $\ket{\Phi'}_{\mathrm{A}}$ [Eq.~\eqref{simplified-loss}] to the ideal state $\ket{\Phi}$ [Eq.~\eqref{send-parity}], these two effects can be expressed in the form of a map $\ket*{C_{\alpha e^{im\varphi}}^\sigma}\ket{\sigma}\mapsto e^{im\varphi}\ket{C_{\alpha e^{im\varphi}}^\sigma}\ket{\bar{\sigma}}$, which, up to the phase factor $e^{im\varphi}$, emphasizes that although the number parity of the cat state is what flipped in actuality, this parity flip ``looks'' like a phase flip $\ket{\pm}\mapsto \ket{\mp}$ of the ancilla. This is what allows it to be detected through the ancilla measurement.

Following the interaction of the light pulse with Bob's qubits [described by Eq.~\eqref{bob-cat-flip}], the state of the light pulse and all qubits is given by
\begin{equation}
    \ket{\Phi_k'}=\frac{1}{\sqrt{2}}R_{\mathrm{A}}^\dagger\left(\ket{\Psi_k}\ket{\Psi_X^+}+\ket{\Psi_{k+2^{N-1}}}\ket{\Psi_X^-}\right),
\end{equation}
where here, $R_{\mathrm{A}}^\dagger$ is the unitary $R$ [Eq.~\eqref{r-operator}] acting on Alice's qubits ($R_{\mathrm{A}}^\dagger\ket{m}_{\mathrm{A}}=e^{im\varphi}\ket{m}_{\mathrm{A}}$). The states $\ket{\Psi_X^\pm}\propto \ket{+-}\pm\ket{-+}$ are once again Bell states of the two ancilla qubits held by Alice and Bob. If undetected, the probabilistic loss of a photon will lead to errors due to the phase factor $e^{im\varphi}$ that is probabilistically ``kicked out'' in front of each $\ket{m}$ [cf.~Eq.~\eqref{simplified-loss}]. In particular, if we assume that this photon is lost with probability $p$, then the final post-measurement state conditioned on heterodyne outcome ${-}k$ will be given by $\rho=(1-p)\ketbra{\Phi_k}+p\ketbra{\Phi_k'}$. However, the parity check of the ancilla qubits can reveal the presence of this unwanted phase while also projecting onto either $\ket{\Phi_k}$ or $\ket{\Phi_{k'}}$, since under the action of Eq.~\eqref{bob-cat-flip}, the parity $\sigma$ of the cat states in Eq.~\eqref{simplified-loss} will be mapped onto the state $\ket{\sigma}$ of Bob's ancilla qubit. Hence, if the parity check returns an eigenvalue of $XX=-1$, the conclusion is that a photon was lost in transit, and that the correction operator $R$ [Eq.~\eqref{r-operator}] should be applied to Alice's qubits to compensate the unwanted phases. Once this correction operator has been applied, the state produced by the independent $Z$-basis measurements of the two ancilla qubits is once again $\ket{\Psi_k}$ for an outcome $\ket{00}$ or $\ket{11}$, and $\ket{\Psi_{k+2^{N-1}}}$ for an outcome $\ket{01}$ or $\ket{10}$. Although we have motivated the correction operator $R$ using a simplified error model that only accounts for single-photon losses, the calculation presented in Appendix \ref{sec:full-loss} makes the connection to the full loss model [Eq.~\eqref{amplitude-damping}, below] explicit. 

\section{Imperfections}\label{sec:errors}

Provided the photon loss probability is sufficiently low, most photon-loss events will be single-photon loss events that can be corrected when using the cat encoding via the procedure given in the previous section. However, it is also possible for multiple photons to be lost in transit, leading to errors that scale like higher powers of the photon-loss probability. Together with the non-orthogonality of the qudit basis states, multiphoton losses are an intrinsic source of error related to the choice of qudit encoding. We quantify these errors in Sec.~\ref{sec:protocol-intrinsic}, below, before considering additional errors due to qubit dephasing in Sec.~\ref{sec:dephasing}. 

\subsection{Protocol-intrinsic errors}\label{sec:protocol-intrinsic}
To quantify errors due to multiphoton loss events (occurring at higher order in the photon-loss probability), we model the effects of photon loss by an amplitude-damping channel~\cite{chuang1997bosonic,bergmann2016quantum}
\begin{equation}\label{amplitude-damping}
    \Lambda(\rho)=\sum_{k=0}^\infty M_k \rho M_k^\dagger,
\end{equation}
where here, $M_k$ is a Kraus operator associated with the loss of $k$ photons, each lost with a probability $1-\eta$:
\begin{equation}\label{krauss-operator}
    M_k=\sqrt{\frac{(1-\eta)^k}{k!}}\sqrt{\eta}^{\hat{a}^\dagger\hat{a}}\hat{a}^k.
\end{equation}
We assume photons are only lost while traveling from Alice's register to Bob's register and neglect all other potential sources of loss. (Intrinsic cavity loss is included in Appendix \ref{sec:entangling-operation}, and we comment on its effect at the end of this section.) By evaluating the action of the loss channel $\Lambda$ on $\ket{\Psi}_{\mathrm{A}}$ [Eq.~\eqref{state-transmitted}] and $\ket{\Phi}_{\mathrm{A}}$ [Eq.~\eqref{send-parity}], we calculate the final state of Alice's and Bob's qubits obtained in the presence of photon loss, neglecting measurement errors in both the heterodyne detection and ancilla-qubit measurements.  For concreteness, we assume that a parity eigenvalue of $XX=+1$ was obtained as part of the cat-qudit protocol. The fidelity $F_\zeta$ of the state $\rho_{\zeta,k}$ produced by the phase- or cat-qudit-mediated protocol ($\zeta=\mathrm{phase},\mathrm{cat}$), relative to the ideal state $\ket{\Psi_k}$ [Eq.~\eqref{ideal-almost-bell}], is given by $F_{\zeta}=\langle \Psi_k\vert \rho_{\zeta,k}\vert \Psi_k\rangle$. The states $\rho_{\zeta,k}$  are derived in Appendix \ref{sec:full-loss} accounting for the full loss channel [Eq.~\eqref{amplitude-damping}], giving
\begin{equation}\label{fidelity-loss}
    F_{\zeta}(n_\ell)=\frac{1}{4^N}\left[2^N+2\sum_{j=1}^{2^N-1}(2^N-j)f_{\zeta}(n_\ell,j)\right],
\end{equation}
where in terms of the average number $n_\ell=(1-\eta)\alpha^2$ of lost photons,
\begin{align}
    &f_{\mathrm{phase}}(n_\ell,j)=e^{-2 n_{\ell}\sin^2{\frac{j\varphi}{2}}}\cos{(n_\ell \sin{j\varphi})},\\
    &f_{\mathrm{cat}}(n_\ell,j)=\cos{(n_\ell\sin{j\varphi})}\frac{\cosh{(n_\ell \cos{j\varphi})}}{\cosh{n_\ell}}.
\end{align}
Summing over $j$, we then find that to leading order in $n_\ell$ and independent of $N$, 
\begin{align}
    &F_{\mathrm{phase}}(n_\ell)=1-n_\ell+O(n_\ell^2),\label{phase-infidelity-scaling}\\
    &F_{\mathrm{cat}}(n_\ell)=1-\frac{n_\ell^2}{2}+O(n_\ell^4).\label{cat-infidelity-scaling}
\end{align}
This result highlights the suppression of errors enabled by the cat-qudit encoding and associated loss detection scheme: Since single-photon losses are detectable, errors due to photon loss are eliminated at first order. [As remarked above, Eq.~\eqref{cat-infidelity-scaling} gives the scaling conditioned on an eigenvalue $XX=+1$. In Appendix \ref{sec:full-loss}, it is shown that for an eigenvalue $XX=-1$, the fidelity is instead given by $F_{\mathrm{cat}}(n_\ell)=1-n_\ell^2/6+O(n_\ell^4)$. The higher fidelity for the same $n_\ell$ in the presence of a heralded photon-loss event ($XX=-1$) can be traced back to the fact that Poisson-distributed photon-loss events have a superexponential dependence on the number of lost photons (cf.~Appendix \ref{sec:full-loss})].

Next, we quantify the measurement errors resulting from the finite distinguishability of different $\ket{k}_{\mathrm{phase}}$ basis states for finite coherent-state amplitudes $\alpha$. The required measurement for both protocols is a projection onto the basis $\{\ket{k}_{\mathrm{phase}}=\ket{\alpha e^{ik\varphi}}\}$. This measurement can be realized using heterodyne detection, which can be modelled by the POVM elements
$\hat{\Pi}(\beta)=\pi^{-1}\ketbra{\beta}$, where here, $\ket{\beta}$ is a coherent state characterized by the complex number $\beta=\beta_1+i\beta_2$ with $\beta_1,\beta_2\in\mathbb{R}$. Given some coherent state $\ket{\alpha}$, the probability density (per unit area in phase space) $p(\beta\vert \alpha)$ of obtaining the measurement outcome $\beta$ is given simply by a coherent-state overlap, $p(\beta\vert \alpha)=\pi^{-1}e^{-\vert \beta-\alpha\rvert^2}$. With maximum-likelihood estimation, a measurement outcome $\beta$ located in the ``wedge'' of phase space bisected by the ray at angle $k\varphi$ and subtended by a total angle of $\varphi$ will be interpreted as a measurement of $\ket*{k}_{\mathrm{phase}}$. Using this maximum likelihood strategy, the probability $p_{\mathrm{m}}(\alpha)$ of making a measurement error when implementing a projection onto the basis  $\{\ket{k}_{\mathrm{phase}}=\ket{\alpha e^{ik\varphi}}\}$ is then given by 
\begin{align}
\begin{aligned}
    p_{\mathrm{m}}(\alpha)
    =1-\frac{1}{\pi}\int_0^\infty d\beta_1 \int_{-\beta_1\tan{\varphi/2}}^{\beta_1\tan{\varphi/2}}d\beta_2 \:p(\beta\vert \alpha),\\
\end{aligned}
\end{align}
where from symmetry considerations, we can express $p_{\mathrm{m}}$ as an integral over the phase-space wedge centered about $\ket{\alpha}=\ket{0}_{\mathrm{phase}}$. Evaluating the integral then gives
\begin{equation}\label{meas-error}
    p_{\mathrm{m}}(\alpha)=\mathrm{erfc}\left(\alpha\sin{\frac{\varphi}{2}}\right).
\end{equation}
Since $\varphi=2\pi/2^N$, achieving a fixed value of $p_{\mathrm{m}}$ as the target number $N$ of Bell pairs is increased therefore requires an increase in the average number $\lvert \alpha\rvert^2$ of photons in the initial coherent state. Reducing measurement errors by increasing $\lvert\alpha\rvert^2$ will, however, come at the cost of increasing errors due to multiphoton losses. This interplay leads to an optimal value of $\alpha$, derived in Sec.~\ref{sec:fidelity-optimization}, below, that maximizes the fidelity of the phase- and cat-qudit-mediated protocols. 

\subsubsection{Fidelity optimization}\label{sec:fidelity-optimization}

If, during heterodyne detection of the light pulse, an outcome $\ket{-k}_{\mathrm{phase}}$ is misidentified as $\ket*{{-}k'}_{\mathrm{phase}}$, then Bob will apply the wrong correction operator 
and produce the state $\ket{\Psi_{k-k'}}$ rather than $\ket{\Psi_0}=\ket{\Phi^+}^{\otimes N}$ as intended. Since $\langle \Psi_k\vert \Psi_{k'}\rangle=\delta_{k,k'}$, a measurement error will result in zero overlap with the target state.  The total error probability $\epsilon_{\zeta}$ of the $\zeta$-qudit-mediated protocol ($\zeta=\mathrm{phase},\mathrm{cat}$) for a fixed average number of photons $n=\lvert \alpha\rvert^2$, transmission probability $\eta$, and target number $N$ of Bell states can therefore be written as
\begin{equation}\label{total-infidelity}
    \epsilon_{\zeta}(N,\eta,n)=1-\left[1-p_{\mathrm{m}}(N,\eta,n)\right]F_{\zeta}(n_\ell).
\end{equation}
Here, $p_{\mathrm{m}}(N,\eta,n)=\mathrm{erfc}(\sqrt{\eta n}\sin{\frac{\pi}{2^N}})$ is the probability of making a measurement error [cf.~Eq.~\eqref{meas-error}] accounting for the attenuation $\alpha\mapsto\sqrt{\eta}\alpha$ of the coherent-state amplitude due to the full loss channel $\Lambda$ [Eq.~\eqref{amplitude-damping}]. The quantity $F_\zeta(n_\ell)$ was given in Eq.~\eqref{fidelity-loss} and sets the fidelity of the state when no measurement error is made. Since the rate of measurement errors decreases with $n$, while the rate of errors due to multiphoton losses increases with $n$, the total error $\epsilon_{\zeta}$ given fixed $N$ and $\eta$ will be minimized for an optimal number $n_{\zeta}=n_\zeta(N,\eta)$ of photons in the initial coherent state. To calculate $n_{\zeta}$, one can set $\frac{\partial \epsilon_{\zeta}}{\partial n}\big\lvert_{n=n_\zeta}=0$, or equivalently,
\begin{equation}\label{nopt}
   \frac{\partial}{\partial n}\mathrm{ln}(1-p_{\mathrm{m}})\bigg\lvert_{n_{\zeta}}=-\frac{\partial}{\partial n}\mathrm{ln}\:F_{\zeta}\bigg\lvert_{n_{\zeta}}.
\end{equation}
Here, we derive closed-form, approximate expressions for $n_{\zeta}(N,\eta)$ valid in the regime 
\begin{align}
\begin{aligned}\label{requirements}
    n_{\ell}=(1-\eta)n<1,\\
    \eta n\: \mathrm{sin}^2(2^{-N}\pi)>1,
\end{aligned}
\end{align}
where the latter condition ensures that measurement errors are small for fixed values of $N$ and $\eta$. (If the probability $1-\eta$ of losing a photon is such that these conditions cannot be met simultaneously for a target $N$, then a heralded single-photon scheme~\cite{lo2019quantum,xie2021quantum,zheng2022entanglement} may be more appropriate as the probability of uncorrectable multiphoton losses increases with $n$. Alternatively, the target $N$ could be decreased.) Together, the requirements given in Eq.~\eqref{requirements} imply that the expressions derived below are valid only for
\begin{equation}\label{regime}
    \Lambda_{N,\eta}\equiv\frac{1-\eta}{\eta\: \sin^2{(2^{-N}\pi)}}<1.
\end{equation}

We now show that, at the optimized value of $n=n_\zeta$, $\epsilon_\zeta$ can be approximated as a single-valued function of $\Lambda_{N,\eta}$. This makes it easier to compare the two protocols for fixed $N,\eta$ while fully accounting for the interplay between error sources, and with both protocols operated at their respective optimal values of $n$. We begin by noting that for an average number of lost photons $n_{\ell}<1$, we can use Eqs.~\eqref{phase-infidelity-scaling}-\eqref{cat-infidelity-scaling} to write
\begin{align}
    &-\frac{\partial}{\partial n}\mathrm{ln}\:F_\mathrm{phase}(n_\ell)= (1-\eta)\left[1+O(n_\ell)\right],\label{f0-lo}\\
    &-\frac{\partial}{\partial n}\mathrm{ln}\:F_\mathrm{cat}(n_\ell)=(1-\eta) \left[n_\ell+O(n_\ell^3)\right].\label{f1-lo}
\end{align}
The left-hand side of Eq.~\eqref{nopt} can meanwhile be expanded as
\begin{align}
    \frac{\partial}{\partial n}\mathrm{ln}(1-p_{\mathrm{m}})=\eta \sin^2{\left(\frac{\pi}{2^N}\right)}\frac{e^{-x}}{\sqrt{\pi x}}\left[1+O\left(\frac{e^{-x}}{\sqrt{x}}\right)\right],\label{pmeas-lo}
\end{align}
where $x=\eta n \sin^2{(2^{-N}\pi)}$. Dropping all subleading corrections in $n_{\ell}$ and $e^{-x}/\sqrt{x}$ in Eqs.~\eqref{f0-lo}, \eqref{f1-lo}, and \eqref{pmeas-lo}, approximate expressions for the optimal average photon number $n_{\zeta}(N,\eta)$ for fixed $N$ and $\eta$ can then be found as solutions to 
\begin{align}
    &\frac{e^{-x_\mathrm{phase}}}{\sqrt{\pi x_\mathrm{phase}}}\simeq  \Lambda_{N,\eta},\label{x0}\\
    &\frac{e^{-x_\mathrm{cat}}}{\sqrt{\pi x_\mathrm{cat}}}\simeq \Lambda_{N,\eta}^2 x_{\mathrm{cat}},\label{x1}
\end{align}
where $x_\zeta=\eta n_\zeta \sin^2{(2^{-N}\pi)}$. Equations \eqref{x0} and \eqref{x1} are approximate reformulations of Eq.~\eqref{nopt} valid for $\Lambda_{N,\eta}<1$. These admit closed-form analytic solutions for the optimal average photon numbers, given by
\begin{align}
    &n_{\mathrm{phase}}(N,\eta)\simeq \frac{1}{2}\frac{\Lambda_{N,\eta}}{1-\eta}W\left(\frac{2}{\pi \Lambda_{N,\eta}^2}\right),\label{n0}\\
    &n_{\mathrm{cat}}(N,\eta)\simeq \frac{3}{2}\frac{\Lambda_{N,\eta}}{1-\eta}W\left(\frac{2}{3}\left[\frac{1}{\sqrt{\pi}\Lambda_{N,\eta}^2}\right]^{2/3}\right),\label{n1}
\end{align}
where here, $W(x)$ is the principal branch of the Lambert-W function. The approximate equalities in Eqs.~\eqref{n0} and \eqref{n1} are inherited from the approximate equalities in Eqs.~\eqref{x0} and \eqref{x1}---no further approximations have been made.

\begin{figure}
    \centering
    \includegraphics[width=\linewidth]{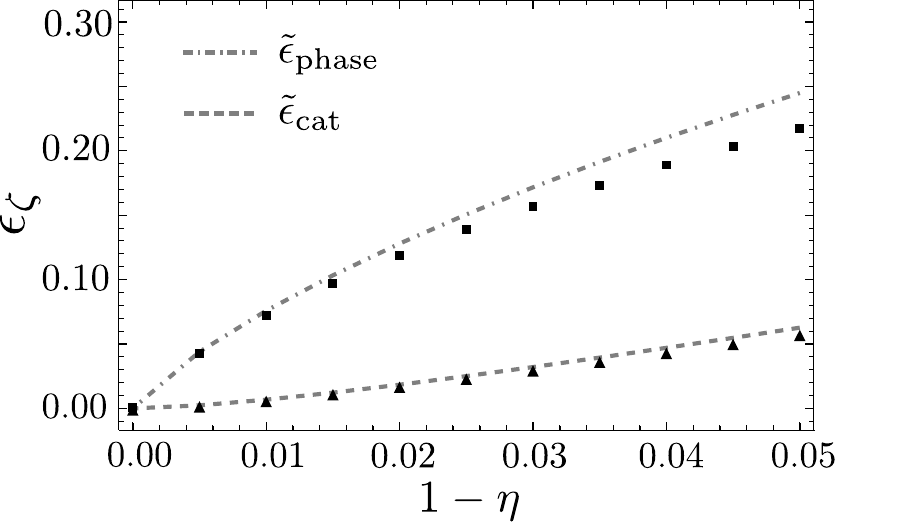}
    \caption{Infidelity $\epsilon_\zeta=\epsilon_\zeta(N,\eta,n_\zeta)$ with $N=2$ as a function of the photon-loss probability $1-\eta$, where here, $n_\zeta$ is an optimized average photon number defined via Eq.~\eqref{nopt}. The black triangles (squares) correspond to a numerical evaluation of Eq.~\eqref{nopt} for $\zeta=\mathrm{cat}$ ($\zeta=\mathrm{phase}$), while the dashed (dot-dashed) line corresponds to the approximate closed-form solution $\Tilde{\epsilon}_\mathrm{cat}$ ($\Tilde{\epsilon}_\mathrm{phase}$) given in Eqs.~\eqref{epsilon-opt}-\eqref{approximate-infidelity}, which is valid only for $\Lambda_{N,\eta}\propto 1-\eta< 1$. }
    \label{fig:errors}
\end{figure}

Next, we use Eqs.~\eqref{x0} and \eqref{x1} to derive approximate expressions (valid for $\Lambda_{N,\eta}<1$) for the total infidelity $\epsilon_\zeta(N,\eta,n)$ evaluated at the optimal $n=n_\zeta$. Since $p_\mathrm{m}\simeq e^{-x}/\sqrt{\pi x}$, we can use Eqs.~\eqref{x0}-\eqref{n1} to express $p_{\mathrm{m}}$ and $n_{\ell}=(1-\eta)n$, both evaluated at $n=n_\zeta$, in terms of $\Lambda_{N,\eta}$ only. Inserting these expressions for $p_{\mathrm{m}}$ and $n_\ell$ back into Eq.~\eqref{total-infidelity}, we then have
\begin{align}
    &\epsilon_\zeta(N,\eta,n_\zeta)\simeq \Tilde{\epsilon}_\zeta(\Lambda_{N,\eta}),\label{epsilon-opt}
\end{align}
where
\begin{align}
\begin{aligned}\label{approximate-infidelity}
    &\Tilde{\epsilon}_\mathrm{phase}(y)=y\left[1+\frac{1}{2}(1-y)W\left(\frac{2}{\pi y^2}\right)\right],\\
    &\Tilde{\epsilon}_\mathrm{cat}(y)=1-\left[1-\frac{3}{2}y^2W\left(h(y)\right)\right]\left[1-\frac{9}{8}y^2 W\left(h(y)\right)^2\right],\\
    &h(y)=\frac{2}{3}\left[\frac{1}{\sqrt{\pi}y}\right]^{2/3}.
\end{aligned}
\end{align}
In the asymptotic limit $\Lambda_{N,\eta}\rightarrow 0$, the Lambert-W function can be approximated by its large-argument asymptotic form, $W(x)\sim \mathrm{ln}\:x$ for $x\gg 1$, and consequently,
\begin{align}
    &\Tilde{\epsilon}_{\mathrm{phase}}(\Lambda_{N,\eta})\sim \frac{\Lambda_{N,\eta}}{2}\mathrm{ln}\left(\frac{2}{\pi \Lambda_{N,\eta}^2}\right),\label{gphase-asymptotic}\\
    &\Tilde{\epsilon}_{\mathrm{cat}}(\Lambda_{N,\eta})\sim \frac{\Lambda_{N,\eta}^2}{2}\mathrm{ln}^2\left(\frac{(2/3)^{3/2}}{\sqrt{\pi}\Lambda_{N,\eta}^2}\right).\label{gcat-asymptotic}
\end{align}
Although $\Tilde{\epsilon}_{\mathrm{phase}}(\Lambda_{N,\eta})>\Tilde{\epsilon}_{\mathrm{cat}}(\Lambda_{N,\eta})$ over the whole interval $0\leq\Lambda_{N,\eta}\lesssim 1$, the asymptotic forms given in Eqs.~\eqref{gphase-asymptotic} and \eqref{gcat-asymptotic} make the first-order suppression of errors enabled by the cat-qudit encoding more readily apparent.  The scaling with $\Lambda_{N,\eta}$ will hold independent of the specific values of $\eta$ and $N$ provided $\Lambda_{N,\eta}<1$.  It should be noted, however, that larger values of $N$ imply more stringent requirements on the channel transmission $\eta$. For instance, a channel with 99\% transmission ($\eta=0.99$) could be used to produce $N=2$ Bell pairs with a fidelity of 99\% using a cat-encoded qudit and 92.5\% using a phase-encoded qudit (Fig.~\ref{fig:errors}). With the same value of $\eta$ but for $N=3$, achievable fidelities are instead 96.5\% and 83.2\% for the cat-encoded and phase-encoded qudit, respectively. 

\begin{table}
\centering
\resizebox{0.47\textwidth}{!}{
\begin{tabular}{c||c|c||c|c}    \toprule
 $N$  & $\eta$ for~$F_{\mathrm{cat}}=99\%$ & $n_\mathrm{cat}$ & $\eta$ for $F_{\mathrm{phase}}=99\%$  & $n_\mathrm{phase}$\\ \toprule
2 & $\;\;$ 98.7\% ($1.3\%$) $\;\;$ & 9.0 & $\;\;$ 99.92\% (0.08\%) $\;\;$  & 10.1 \\
3 & 99.6\% (0.4\%) & 30.2 & 99.98\% (0.02\%) & 35.5 \\
4 & 99.9\% (0.1\%) & 117.3 & 99.99\% (0.01\%) & 121.0\\\bottomrule
\end{tabular}}
\caption{Values of the channel transmission $\eta$, expressed as a percentage, required for 99\% state fidelity when using the cat- or phase-qudit mediated protocol to produce a target number $N$ of Bell pairs in a single shot. For convenience, we indicate in brackets the associated photon-loss probabilities $1-\eta$ with the goal of emphasizing the order-of-magnitude difference in channel-quality requirements. We also give the optimal average number of photons $n_\mathrm{cat}$ and $n_\mathrm{phase}$ associated with these values of $\eta$ and $N$ [Eqs.~\eqref{n0}-\eqref{n1}]. }
\label{tab:targetfidelities}
\end{table}

Instead of fixing $\eta$ and determining the achievable state fidelities for different $N$, we can also fix a target fidelity and calculate combinations of $(N,\eta)$  capable of achieving that target fidelity. We then observe that although the requirements on $\eta$ become more stringent with increasing $N$ for both protocols, the ability to flag single-photon losses and correct their backaction translates into less stringent requirements for $\eta$ \textit{at fixed $N$} for $\zeta=\mathrm{cat}$ relative to $\zeta=\mathrm{phase}$. For instance, achieving a target fidelity of 99\% [i.e.~$\epsilon_\zeta(\Lambda_{N,\eta})=0.01$] requires that $\Lambda_{N,\eta}\approx 0.0017$ for $\zeta=\mathrm{phase}$ and $\Lambda_{N,\eta}\approx 0.027$ for $\zeta=\mathrm{cat}$. By fixing $N$, these required values of $\Lambda_{N,\eta}$ can be translated into requirements on the channel transmission $\eta$ (Table \ref{tab:targetfidelities}). Notably, Table \ref{tab:targetfidelities} clearly illustrates that for the values of $N$ taken here, the ability to correct single-photon losses leads to roughly an order-of-magnitude reduction in the channel quality needed to achieve 99\% fidelity. Photon loss rates of $0.1\%$/m have been measured experimentally for propagation in microwave-frequency waveguides~\cite{kurpiers2017characterizing}. However, given the dependence of $\Lambda_{N,\eta}$ on $N$, it must be acknowledged that the protocols presented here are better suited to smaller values of $N$, and that $N$ cannot be arbitrarily increased in any realistic implementation. 

Although the cat-qudit encoding is designed to protect against photon loss from the channel connecting Alice and Bob, intrinsic cavity losses (with rate $\kappa_\mathrm{int}$) to some uncontrolled environment will also lead to imperfections in the entangling operation itself. These errors can be quantified using the expression for the reflection coefficient given in Eq.~\eqref{reflection-coefficient} of Appendix~\ref{sec:entangling-operation}. In particular, errors in the phase $\varphi_j$ scale like $(\chi_j/\kappa_j) (\kappa_{\mathrm{int}}/\kappa_j)^2$ for $\kappa_{\mathrm{int}}<\kappa_j$. In addition to phase errors, intrinsic losses will also induce a reduction in amplitude of the light pulse, which contributes errors $\sim \kappa_{\mathrm{int}}/\kappa_j$ for $\kappa_\mathrm{int}<\kappa_j$. For a coherent-state amplitude $\alpha$, this implies that the probability of losing a photon from inside the cavity during the entangling operation is given roughly by $(\kappa_{\mathrm{int}}/\kappa_j)\vert\alpha\rvert^2$. Hence, in a near-term implementation where $\kappa_{\mathrm{int}}$ represents a significant fraction of $\kappa_j$, intrinsic losses could also be the limiting error source when attempting to increase $N$.

\subsection{Qubit dephasing}\label{sec:dephasing}

Errors may also arise due to dephasing of the stationary qubits resulting from a shot-to-shot variation of the qubit splitting. For phase shifts $\varphi_j$ realized through dispersive cavity-qubit coupling, we show in Appendix \ref{sec:dephasing-appendix} that a variation $\delta\omega_j$ in the frequency $\omega_j\rightarrow \omega_j+\delta\omega_j$ of Alice's $j^{\mathrm{th}}$ qubit will lead to a variation $\delta\varphi_j$ in the phase $\varphi_j$ imprinted on the light pulse, which, for $\lvert \delta\omega_j\rvert<\lvert\Delta_j\rvert$, is given by 
\begin{equation}\label{delta-phi}
    \delta\varphi_j=-2\sin{\left(\frac{\varphi_j}{2}\right)}\frac{\delta\omega_j}{\Delta_j},
\end{equation}
where here, $\Delta_j=\omega_j-\omega_{\mathrm{c}}$ is the detuning between the $j^{\mathrm{th}}$ qubit and its cavity (whose frequency we denote $\omega_{\mathrm{c}}$). The analogous $\delta\varphi_j$ for Bob's $j^{\mathrm{th}}$ qubit can be obtained from Eq.~\eqref{delta-phi} by sending $\varphi_j\rightarrow -\varphi_j$, but with a fluctuation $\delta\omega_j$ that is independent of that for Alice's $j^{\mathrm{th}}$ qubit. For concreteness and simplicity, we assume in this section that the ancilla qubits are coupled to their own separate cavities, in which case the susceptibility of the ancilla-conditioned phase shifts to random variations in the ancilla-qubit frequencies is also described by Eq.~\eqref{delta-phi} with $\varphi_j\rightarrow \pi$.

Variations in the qubit splitting will consequently lead to errors via two distinct mechanisms: In addition to the usual dephasing between different qubit states, there is also shot-to-shot variation in the qubit-state-dependent phase shifts $\varphi_j$ imprinted on the light pulse as part of the entangling dynamics. To quantify these errors, we calculate the average fidelities $\mathcal{F}_{\mathrm{phase}}(t)=\llangle \lvert \langle \Psi\vert\Psi_{\delta\omega}(t)\rangle\rvert^2\rrangle$ and $\mathcal{F}_{\mathrm{cat}}(t)=\llangle \lvert \langle \Phi\vert\Phi_{\delta\omega}(t)\rangle\rvert^2\rrangle$ of the states $\ket{\Psi_{\delta\omega}(t)}$ and $\ket{\Phi_{\delta\omega}(t)}$ obtained in the presence of qubit-frequency variations, relative to the ideal states $\ket{\Psi}$ [Eq.~\eqref{pre-meas-state}] and $\ket{\Phi}$ [Eq.~\eqref{final-cat-2}].
Here, we denote by $\llangle\rrangle$ an average over the probability distribution governing the qubit-frequency variations. In Appendix \ref{sec:dephasing-appendix}, we show that for zero-mean Gaussian-distributed qubit-frequency fluctuations (taken to be uncorrelated across different qubits),
\begin{align}
    \mathcal{F}_{\mathrm{phase}}(t)&=G(t),\label{dephasing-phase}\\
    \mathcal{F}_{\mathrm{cat}}(t)&=G
    (t)G_{\mathrm{a}}(t),
\end{align}
where $G(t)$ and $G_{\mathrm{a}}(t)$ quantify errors due to dephasing of Alice's and Bob's $2N$ register qubits [$G(t)$] and ancilla qubits [$G_{\mathrm{a}}(t)$], respectively: 
\begin{align}
    G(t)&=\frac{e^{-\lvert\alpha\rvert^2\Theta^2}}{4^N}\prod_{\mathcal{X}=\mathrm{A,B}}\sum_{k=0}^N\sum_{\mathcal{S}\in\mathcal{S}_k}e^{-\frac{1}{T_2^{*2}}\sum\limits_{j\in\mathcal{S}}\mathcal{X}_j^2(t)},\label{g(t)}\\
    G_{\mathrm{a}}(t)&=\frac{e^{-\lvert\alpha\rvert^2\Theta_{\mathrm{a}}^2}}{16}\prod_{\mathcal{X}=\mathrm{A,B}}\left[2+2e^{-\frac{\mathcal{X}_{\mathrm{a}}^2(t)}{T_2^{*2}}}\right].\label{ga(t)}
\end{align}
Here, $T_2^*$ is a dephasing time that we take to be the same for all qubits, and which is related to qubit-frequency variations $\delta\omega_{j}$ via $\llangle\delta\omega_{j}^2\rrangle=2/T_2^{*2}$. In Eq.~\eqref{g(t)}, the sum over $\mathcal{S}_k$ is a sum over the set of sets of cardinality $k$ consisting of all distinct sets of $k$ qubits. (For each value of $k$, there are $\binom{N}{k}=N!/[k!(N-k)!]$ such sets, giving $\sum_k\binom{N}{k}=2^N$ terms in the full sum over $k$. As an example, for $N=3$ and $k=2$, $\mathcal{S}_2$ is the set of all possible pairs of qubits: $\mathcal{S}_2=\{\{0,1\},\{0,2\},\{1,2\}\}$.) This sum over sets of fixed cardinality reflects the fact that qubit dephasing is more damaging to coherences between computational basis states $\ket{m}_{\mathrm{A}}\ket{n}_{\mathrm{B}}$ whose binary representations are separated by larger Hamming distances. Finally, the quantities $\Theta^2$, $\Theta_{\mathrm{a}}^2$, $\mathcal{X}_j(t)$, and $\mathcal{X}_{\mathrm{a}}(t)$ appearing in Eqs.~\eqref{g(t)} and \eqref{ga(t)} are given by
\begin{align}
    &\Theta^2=\sum_{j=0}^{N-1}\sum_{\mathcal{X}=\mathrm{A,B}}\sin^2{\left(\frac{\varphi_j}{2}\right)}\frac{1}{(T_2^{*}\Delta_{\mathcal{X},j})^2},\\
    &\Theta_{\mathrm{a}}^2=\sum_{\mathcal{X}=\mathrm{A,B}}\frac{1}{(T_2^{*}\Delta_{\mathcal{X},\mathrm{a}})^2},\\
    &\mathcal{X}_j(t)=t-2\sin{\left(\frac{\varphi_j}{2}\right)}\frac{\lvert\alpha\rvert^2}{\Delta_{\mathcal{X},j}},\\
    &\mathcal{X}_{\mathrm{a}}(t)=t-\frac{2\lvert\alpha\rvert^2}{\Delta_{\mathcal{X},\mathrm{a}}}.\label{X-ancilla}
\end{align}
Here, we indicate by $\mathcal{X}=\mathrm{A,B}$ that $\Delta_{\mathcal{X},j}$ is the detuning of Alice's or Bob's $j^{\mathrm{th}}$ qubit, and similarly for Alice's or Bob's ancilla ($\Delta_{\mathcal{X},\mathrm{a}}$). 

Equations~\eqref{dephasing-phase}-\eqref{X-ancilla} can be used to quantify the errors introduced by qubit dephasing over a total time $t=T$. To identify the regime where these errors are small, we now set the magnitude of all detunings to the same value $\Delta$, and we derive simple conditions controlling the infidelities $1-\mathcal{F}_\zeta$ ($\zeta=\mathrm{phase},\,\mathrm{cat}$), at leading order in $1/T_2^*$. In general, $T$ can be expected to scale like $T\sim N_{\mathrm{tot}}/\kappa+T_{\mathrm{travel}}$ for $N_{\mathrm{tot}}$ cavities, where $\kappa$ is the typical size of the cavity decay rates and $T_{\mathrm{travel}}$ is the time required for light to travel between Alice's and Bob's registers. Depending on the implementation, one term or the other may be dominant, and in the following, we take $T\sim N_{\mathrm{tot}}/\kappa$. This is a reasonable assumption for MHz-scale cavity decay rates and travel times on the order of nanoseconds or tens of nanoseconds, as would be realistic for meter-scale-separated circuit QED devices. 

After expanding $G(t)$ [Eq.~\eqref{g(t)}] to leading order in $1/T_2^*$, we identify the relevant dimensionless expansion parameters as $\lvert \alpha\rvert^2\Theta^2 \ll 1$ and 
\begin{equation}\label{explanation}
    \frac{1}{2^N}\sum_{k=0}^N \sum_{\mathcal{S}\in \mathcal{S}_k}\frac{1}{T_2^{*2}}\sum_{j\in\mathcal{S}}\mathcal{X}_j^2(t)=\frac{1}{2T_2^{*2}}\sum_{j=0}^{N-1}\mathcal{X}_j^2(t)\ll 1,
\end{equation}
where the first equality follows from the fact that each $j$ appears $2^{N-1}$ times in the sum: $\sum_k\sum_{\mathcal{S}\in\mathcal{S}_k}\sum_{j\in\mathcal{S}}\mathcal{X}_j^2=2^{N-1}\sum_j \mathcal{X}_j^2$.
From Eq.~\eqref{explanation}, a small infidelity requires that  $\sum_j(N_{\mathrm{tot}}/\kappa-2\sin{(\varphi_j/2)}\lvert\alpha\rvert^2/\Delta)^2\ll T_2^{*2}$. This condition can be satisfied by requiring that each term in the sum be $\ll T_2^{*2}/N$, which can be satisfied (conservatively) under the conditions
\begin{align}\label{regime-dephasing}
    \frac{N_{\mathrm{tot}}\sqrt{N}}{\kappa T_2^*},\frac{2\lvert\alpha\rvert^2\sqrt{N}}{\Delta T_2^*}\ll 1.
\end{align}
Satisfying the requirement $2\lvert \alpha\rvert^2/\Delta\ll T_2^*/\sqrt{N}$ would already imply that $\lvert \alpha\rvert^2\Theta^2\ll1$ (the other requirement stated at the beginning of this paragraph). In principle, the additional suppression of $\mathcal{F}_{\mathrm{cat}}(t)$ due to $G_\mathrm{a}(t)$ could lead to larger infidelities for the cat-qudit versus phase-qudit protocol, despite the additional protection against photon loss provided by the qudit encoding. However, additional errors due to the presence of the ancilla qubits will typically be smaller than those coming from $G(t)$ since $G_\mathrm{a}(t)$ does not involve any sums over $j=0,\dots,N-1$. Hence, the main impact of including the ancilla qubits is whether $N_{\mathrm{tot}}=2N$ or $2(N+1)$ in Eq.~\eqref{regime-dephasing}.

The condition $\sqrt{N}\lvert\alpha\rvert^2/(\Delta T_2^*)\ll 1$  in Eq.~\eqref{regime-dephasing} can be understood as controlling the size of errors introduced by variations $\delta\varphi_j$ in the phase shifts $\varphi_j$ imprinted on the light pulse, which, as expected, grow with the average number of photons $\lvert\alpha\rvert^2$. These errors can be suppressed for fixed $T_2^*$ by operating with larger detunings $\Delta$. The parameter $N_{\mathrm{tot}}\sqrt{N}/(\kappa T_2^*)$, meanwhile, controls the size of errors due to dephasing between different computational basis states. The scaling with $N_{\mathrm{tot}}\sqrt{N}$ can be understood as follows: In the presence of quasistatic noise, the error due to dephasing over a time $T$ is expected to scale like $2N(T/T_2^*)^2$ for $2N$ qubits with uncorrelated random frequency shifts and for $T<T_2^*$. Since $T\sim N_{\mathrm{tot}}/\kappa$, the overall scaling goes like $N_{\mathrm{tot}}^2N$, and errors can be suppressed by requiring that $N_{\mathrm{tot}}\sqrt{N}/(\kappa T_2^*)\ll 1$ as reported in Eq.~\eqref{regime-dephasing}. For a fixed $T_2^*$, these errors $\propto (\kappa T_2^*)^{-1}$ could be reduced by using leakier cavities (i.e.~larger $\kappa$). The quantities $\Delta$ and $\kappa$ cannot be increased without bound, however, as realizing a certain value of $\varphi_j$ requires that $\chi_j$ and $\kappa_j$ satisfy $\varphi_j=4\arctan{(2\chi_j/\kappa_j)}$ (Appendix \ref{sec:entangling-operation}). Here, $\chi_j=g_j^2/\Delta_j$, where $g_j$ is the strength of the transverse coupling between the $j^{\mathrm{th}}$ qubit and its cavity. Increasing $\Delta_j$ and $\kappa_j$ with the goal of reducing errors due to qubit dephasing therefore requires a larger value of $g_j$ to maintain a fixed $\varphi_j$.  Since $\varphi_j\leq \varphi_{N-1}=\pi$ for all qubits $j$ [Eq.~\eqref{phase-breakdown}], the dispersive shifts required for this protocol all satisfy $\chi_j \leq \kappa/2$. For $\kappa/2\pi=50$ MHz and $g/2\pi=250$ MHz, a value of $\chi=\kappa/2$ can be achieved with a detuning of $\Delta/2\pi=2.5$ GHz. For dephasing on a timescale $T_2^*=10\:\mu$s, we then find the order-of-magnitude $(\Delta T_2^*)^{-1}\sim 10^{-6}$, limiting the average number of photons to $\lvert \alpha\rvert^2\lesssim 10^5$ for values of $N<10$ according to Eq.~\eqref{regime-dephasing}. A dephasing time of $T_2^*=10\:\mu$s could be realized with both spin qubits and superconducting qubits~\cite{kjaergaard2020superconducting,stano2022review}. Whether the limitation on $\lvert \alpha\rvert^2$ is more or less restrictive than the limitation due to multiphoton losses will depend on the value of the channel transmission $\eta$ as discussed in Sec.~\ref{sec:fidelity-optimization}. However, it will  not be the principal limitation on $\lvert\alpha\rvert^2$ if we assume a photon-loss probability $1-\eta$ on the order of 1\%, which translates to optimal photon numbers on the order of $n_\zeta \sim 10^1{-}10^2$ for $N=2,3,4$ [cf.~Eq.~\eqref{n0}-\eqref{n1}]. 
Meanwhile, for the same values of $T_2^*=10\:\mu$s and $\kappa/2\pi=50$ MHz, we have $N_{\mathrm{tot}}\sqrt{N}/(\kappa T_2^*)< 0.01$ for $N<5$ for the cat-qudit protocol and $N<6$ for the phase-qudit protocol.

\section{Discussion and Conclusion}\label{sec:conclusions}

The generation of entanglement between distantly separated groups of qubits, ``Alice'' and ``Bob'', has applications in both quantum communication and distributed quantum computing. The rate at which entanglement is generated can be increased relative to strategies mediated by a single photonic qubit by adopting strategies for parallelized Bell-state generation involving the transmission of a higher-dimensional photonic qudit. Rather than encode the qudit in the state of a single photon using, e.g., a time-bin basis, we have presented in this paper a strategy that could be used to generate $N$ Bell pairs with a hybrid light-matter encoding that leverages the phase degree-of-freedom of a propagating coherent state. The loss of a photon from the light pulse can be detected through a parity measurement involving the matter portion of the qudit, and its backaction subsequently corrected. 

We have analyzed the dominant error sources intrinsic to the choice of qudit encoding, together with errors due to qubit dephasing, and we showed that the ability to correct single-photon loss suppresses errors (accounting for an interplay with measurement errors) to leading order in $\Lambda_{N,\eta}=(1-\eta)/[\eta \sin^2{(\pi/2^N)}]$, provided the photon-loss probability $1-\eta$ is low enough to achieve $\Lambda_{N,\eta}<1$.  Since the requirements on $1-\eta$ become increasingly stringent as the target number $N$ of Bell pairs is increased, it should again be acknowledged that the strategies proposed here may be better suited to smaller $N$ in the near term.  If $1-\eta$ does not meet the requirements even for $N= 1$, then other existing protocols may be more appropriate: For instance, a single Bell state could be created using a higher-order cat qubit enabling multiphoton loss events to be detected~\cite{li2023memoryless}, at the cost of introducing more ancillas, or alternatively, multiple Bell pairs could be created using a time-bin qudit encoded in the state of a single photon~\cite{xie2021quantum,zheng2022entanglement}, allowing the transmission of the qudit to be heralded when successful. In microwave platforms, the loss per unit distance has been found to be as low as 0.005 dB/m~\cite{kurpiers2017characterizing}, corresponding to a photon loss rate of $1-\eta\approx 0.00115$ per meter. By contrast, chiral elements like circulators tend to produce higher losses than free propagation due to circulators typically being off chip (at a higher-temperature stage of the dilution refrigerator); in the near term, the development of on-chip circulators~\cite{zhang2021charge,navarathna2023passive} will consequently address one of the main sources of photon loss affecting current devices. In certain settings, limiting $N$ may also be necessary due to the Boolean operation $\ket{m}\rightarrow \ket{m\oplus -k}$ that Bob must apply to his qubits, conditioned on the outcome $k$ of the heterodyne measurement, in order to recover a product of $N$ Bell pairs $\ket{00}+\ket{11}$ shared between him and Alice. Since this operation involves a quantum Fourier transform, the depth of the circuit required for its implementation will grow as $N$ is increased, introducing more opportunities for errors due to, e.g., qubit decoherence and imperfect gate fidelities.

Going beyond qudit-specific errors, we have also presented an analysis of errors due to qubit dephasing. Beyond increasing the cavity decay rate and qubit detuning, as discussed in Sec.~\ref{sec:dephasing}, other strategies for reducing the impact of qubit dephasing could also be devised. For instance, we assumed that all qubits were prepared in $\ket{+}$ at time $t=0$. In this scenario, each qubit dephases over the total time $T$ required to execute the protocol. However, in an actual implementation, qubit initialization could be staggered to coincide with the arrival of the light pulse, so that qubits interacting with the light pulse at later times have less time to dephase overall. In addition, the results presented in this work were derived under the assumption that all qubits dephase under the influence of free-induction decay. The timescale for qubit dephasing could potentially be increased by subjecting idle qubits---those not interacting with the light pulse at a given point in time---to dynamical decoupling sequences designed to mitigate the impact of noise. However, any dynamical decoupling should be performed with the qubit far detuned from its cavity. This will prevent qubit coherence from being converted into cavity coherence by residual transverse interactions, resulting in unintended photon emission~\cite{mcintyre2022non}. 

The benefits of creating several Bell pairs at once are clear in a quantum-communication context where $T_{\mathrm{travel}}$ constitutes a sizeable fraction of the total protocol time $T$. Hence, a final remark may be in order concerning the circuit-QED-inspired scenario analyzed in Sec.~\ref{sec:dephasing}, in which qubits dephase over a total time $T\sim N_{\mathrm{tot}}/\kappa$, scaling with $N_{\mathrm{tot}}$ and with a negligible contribution coming from $T_{\mathrm{travel}}$. From the standpoint of qubit dephasing, $N_{\mathrm{tot}}$-linear scaling may present advantages relative to schemes involving time-bin qudits where $T\propto 2^{N_{\mathrm{tot}}}$~\cite{zheng2022entanglement}. However, the scaling of $T$ with $N_{\mathrm{tot}}$ will lead to comparable amounts of dephasing as one would expect when creating $N$ Bell states \textit{sequentially} using similar qubit-state-conditioned phase shifts~\cite{mcintyre2024flying}. In a circuit QED context, the principal advantage of the parallelized scheme is that the channel connecting Alice and Bob is only occupied for a duration $\sim \kappa^{-1}$ set by the pulse duration (which is related by the speed of light to the spatial extent of the quasimode supporting the coherent state $\ket{\alpha}$), independent of $N$. In particular, the light pulse is localized in either Alice's or Bob's register for the entire duration of the protocol apart from the time required for it to traverse the channel, leaving the channel free to be used for other purposes at intermediate times.  In a sequential scheme, the channel will instead be occupied for a duration $N\kappa^{-1}$ as \textit{each} Bell state requires that a light pulse be sent from Alice to Bob. The parallelized scheme presented here therefore provides greater flexibility for scheduling and synchronizing operations across the whole device.

\begin{acknowledgments}
We acknowledge funding from the Natural Sciences and Engineering Research Council of Canada (NSERC) and from the Fonds de recherche du Qu\'ebec--Nature et technologies (FRQNT). 
\end{acknowledgments}

\appendix

\section{Entangling operation}\label{sec:entangling-operation}

In this Appendix, we first describe how the entangling operation given in Eq.~\eqref{phase-shifts} can be implemented in practice. We then describe how an ancilla qubit could be incorporated to realize the entangling operation given in Eq.~\eqref{alice-entangle-cat}.

For a state $\ket{m}$ of Alice's qubits, a phase shift $\ket{\alpha}\mapsto\ket*{e^{im\varphi}\alpha}$ could be realized by successively reflecting a coherent light pulse (prepared in state $\ket{\alpha}$) off a series of $N$ single-sided cavities each containing one of Alice's $N$ qubits (Fig.~\ref{fig:setup}). For concreteness, we consider a situation where qubit $j$ is dispersively coupled with strength $\chi_j$ to its cavity according to the Hamiltonian 
\begin{equation}
    H_j=\chi_j \sigma_{zj} a_j^\dagger a_j,
\end{equation}
where $\sigma_{zj}=\ketbra{0}_j-\ketbra{1}_j$ is a Pauli-Z operator acting on qubit $j$ and $a_j$ is an annihilation operator that removes one photon from cavity $j$. For a single-sided cavity with decay rate $\kappa_j$ to the external transmission line, the quantum Langevin equation describing the dynamics of the cavity field in the presence of driving by an input field $r_{\mathrm{in}}(t)$ (in this case, the coherent light pulse) is given by
\begin{equation}\label{langevin-eq}
    \langle\dot{a}_j\rangle_t=-\frac{1}{2}(\kappa_j+\kappa_{\mathrm{int}})\langle a_j \rangle_t -i(-1)^{s_j}\chi_j\langle a_j\rangle_t-\sqrt{\kappa_j}r_{\mathrm{in}}(t),
\end{equation}
where $s_j=0$ ($1$) for qubit $j$ in state $\ket{0}$ ($\ket{1}$), and where $\kappa_{\mathrm{int}}$ accounts for the effects of intrinsic cavity losses. Combined with the input-output relation $r_{\mathrm{out}}(t)=r_{\mathrm{in}}(t)+\sqrt{\kappa_j}\langle a_j\rangle_t$~\cite{gardiner1985input}, Eq.~\eqref{langevin-eq} can be used to calculate the reflection coefficient $R_{s_j}(\omega)=r_{\mathrm{out}}(\omega)/r_{\mathrm{in}}(\omega)$ describing the phase acquired by the input field upon reflection from the cavity, conditioned on $s_j=0,1$: 
\begin{equation}\label{reflection-coefficient}
    R_{s_j}(\omega)=\frac{i\left(\omega+(-1)^{s_j}\chi_j \right)-\frac{1}{2}(\kappa_j-\kappa_{\mathrm{int}})}{i\left(\omega+(-1)^{s_j}\chi_j \right)+\frac{1}{2}(\kappa_j+\kappa_{\mathrm{int}})}.
\end{equation}
Defining $\theta_{s_j}=\mathrm{arg}\:R_{s_j}(0)$ to be the phase shift acquired when driving at the bare cavity frequency ($\omega=0$ in this frame), we then find that for $\chi_j\neq 0$ and $\kappa_{\mathrm{int}}=0$,
\begin{align}
    \theta_{s_j}&=(-1)^{s_j} \left[\mathrm{sgn}(\chi_j)\pi-2\arctan{\frac{\chi_j}{\kappa_j/2}}\right],
\end{align} 
where $\mathrm{sgn}(\chi)=\pm 1$ is the sign of $\chi$. (In the case where $\chi_j=0$, $\theta_{s_j}=\pi$ independent of $s_j$.) Since $\theta_{0}=-\theta_{1}$, the phase shift acquired conditioned on state $\ket{0}$ of qubit $j$ cannot be set to zero as assumed in Eq.~\eqref{phase-shifts}. However, a phase \textit{difference} of 
\begin{equation}
    \varphi_j=\theta_{s_j=1}-\theta_{s_j=0}\;(\mathrm{mod}\:2\pi)
\end{equation} can be realized by choosing  
\begin{equation}\label{dispersive-shifts}
    \chi_j=\frac{\kappa_j}{2}\tan{\frac{\varphi_j}{4}}.
\end{equation}
Since $\varphi\leq \varphi_j\leq \pi$ [Eq.~\eqref{phase-breakdown}], we have $\chi_j\leq \kappa_j/2$ for all $j$. To implement the entangling operation given in Eq.~\eqref{phase-shifts}, the phase reference of $\alpha$ can simply be redefined to compensate for the phases acquired conditioned on the qubits being in their $\ket{0}$ states: $\alpha\mapsto \alpha \prod_j e^{-i\theta_{s_j=0}}$. 

To account for errors due to a finite $\kappa_{\mathrm{int}}$, we can expand both $\theta_{s_j}=\mathrm{arg}\:R_{s_j}(0)$ and $\lvert R_{s_j}(0)\vert^2$ to leading order in $x=\kappa_{\mathrm{int}}/\kappa_j$:
\begin{align}
    &\theta_{s_j}(x)=\theta_{s_j}(x=0)-\frac{4(-1)^{s_j}\chi_j\kappa_j^3x^2}{(\kappa_j^2+4\chi_j^2)^2}+O(x^4),\label{phase-kappaint}\\
    &\vert R(0)\rvert^2=1-\frac{4\kappa_j^2x}{\kappa_j^2+4\chi^2}+O(x^3).\label{reflection-kappaint}
\end{align}
In the case where $\kappa_j/2\gg \chi_j$, the corrections are $\sim (\chi_j/\kappa_j) x^2=(\chi_j/\kappa_j)(\kappa_{\mathrm{int}}/\kappa_j)^2$ for the phase $\theta_{s_j}$ and $\sim \kappa_{\mathrm{int}}/\kappa_j$ for the reflection probability, as reported in the main text. In the event that $\kappa_j/2$ and $\chi_j$ are comparable, the corrections can be read off directly from Eqs.~\eqref{phase-kappaint} and \eqref{reflection-kappaint}. 


If the ancilla qubit is coupled to its own cavity, then Eq.~\eqref{dispersive-shifts} can be used to set the dispersive coupling of the ancilla as well. However, we now explain how the entangling operation required for the cat-qudit protocol [cf.~Eq.~\eqref{alice-entangle-cat}] could be realized with dispersive coupling of two qubits to a common cavity mode. To derive the required dispersive couplings for this setup, we consider a cavity with decay rate $\kappa$ coupled to two qubits, indexed 1 and 2 for simplicity. We assume that the two qubits are detuned from the cavity by the same amount, but with detunings of opposite sign. This allows us to neglect a virtual-photon-mediated qubit-qubit coupling of the form $J(\sigma_+^{(1)}\sigma_-^{(2)}+\mathrm{h.c.})$ entering the Hamiltonian at the same order in $g_{1,2}$, for which $J=0$ under the assumptions laid out above~\cite{blais2007quantum,lalumiere2010tunable}. In particular, this also implies that the dispersive couplings $\chi_1$ and $\chi_2$ of the two qubits will be opposite in sign. 

By a similar calculation to that performed above, the reflection coefficient $R_{s_{1}s_{2}}(\omega)$ conditioned on state $\ket{s_1s_2}$ is found to be 
\begin{equation}\label{reflection-coeff-2}
    R_{s_{1}s_{2}}(\omega)=\frac{i\left(\omega+\sum\limits_{j=1,2}(-1)^{s_j}\chi_j\right)-\frac{\kappa}{2}}{i\left(\omega+\sum\limits_{j=1,2}(-1)^{s_j}\chi_j\right)+\frac{\kappa}{2}}.
\end{equation} 
With $\theta_{s_1s_2}=\mathrm{arg}\:R_{s_1s_2}(0)$, we then have
\begin{align}
    &\theta_{00}=-\theta_{11}=\mathrm{sgn}(\chi_+)\pi-2\:\mathrm{arctan}\frac{2\chi_+}{\kappa},\label{theta_00}\\
    &\theta_{01}=-\theta_{10}=\mathrm{sgn}(\chi_-)\pi-2\:\mathrm{arctan}\frac{2\chi_-}{\kappa},\label{theta_11}
\end{align}
where $\chi_\pm=\chi_1\pm \chi_2$. In an ideal scenario, reflection from the cavity would realize an entangling operation given by
\begin{equation}\label{app:entangling}
    \ket{s_1s_2}\ket{\alpha}\mapsto\ket{s_1s_2}\ket*{\alpha e^{i(\pi s_1+\varphi_j s_2)}}
\end{equation}
for some choice of $\varphi_j$ [defined in Eq.~\eqref{phase-breakdown}]. [Qubit 2 whose state $\ket{s_2}$ appears in Eq.~\eqref{app:entangling} would then correspond to qubit $j$ in the context of Alice's full register.] From Eq.~\eqref{app:entangling}, we see that the two-qubit-conditioned phase shifts $\theta_{s_1s_2}$ are ideally $\theta_{00}=0$, $\theta_{01}=\varphi_j$, $\theta_{10}=\pi$, and $\theta_{11}=\pi+\varphi_j$. However, since Eq.~\eqref{reflection-coeff-2} imposes the constraints $\theta_{00}=-\theta_{11}$ and $\theta_{01}=-\theta_{10}$, we can only realize the operation given in Eq.~\eqref{app:entangling} up to a global phase-space rotation of $\phi$, giving
\begin{align}
\begin{aligned}\label{phases}
    \theta_{00}&=\phi,\\
    \theta_{01}&=\phi+\varphi_j,\\
    \theta_{10}&=\phi+\pi,\\
    \theta_{11}&=\phi+\pi+\varphi_j.
\end{aligned}
\end{align}
From the constraints $\theta_{00}=-\theta_{11}$ and $\theta_{10}=-\theta_{01}$, we obtain a global phase of 
\begin{equation}\label{global-phase}
    \phi=-\frac{\pi+\varphi_j}{2}.
\end{equation}
To obtain Eq.~\eqref{app:entangling}, the phase shift of $\phi$ can be compensated after the entangling operation by a redefinition of the phase reference $\alpha\mapsto\alpha e^{-i\phi}$ to recover $\theta_{00}=0$. 

Equations \eqref{theta_00}, \eqref{theta_11}, \eqref{phases}, and \eqref{global-phase} give two equations in two unknowns $\chi_\pm$, in terms of which we can find the dispersive shifts $\chi_{1,2}=(\chi_+\pm\chi_-)/2$ required to realize Eq.~\eqref{app:entangling} (up to a global phase of $\phi$). Although Eqs.~\eqref{theta_00} and \eqref{theta_11} involve the sign of $\chi_\pm$, all four sign combinations give the same result for $\chi_{1,2}$:
\begin{align}
    \chi_1&=-\frac{\kappa}{2}\sec{\left(\frac{\varphi_j}{2}\right)},\label{chi-a}\\
    \chi_2&=\frac{\kappa}{2}\tan{\left(\frac{\varphi_j}{2}\right)}.\label{chi-data}
\end{align}
The dispersive shifts so obtained have opposite signs, as required to eliminate the spurious cavity-mediated coupling $J$. Since $\sec{(x)}$ is an even function of $x$ while $\tan{(x)}$ is odd, we cannot send $\varphi_j\rightarrow -\varphi_j$ in Eqs.~\eqref{chi-a}-\eqref{chi-data} as a way of obtaining Bob's dispersive couplings without losing the property that $\mathrm{sgn}(\chi_1)=-\mathrm{sgn}(\chi_2)$. (Recall that for Bob, the phase shift $\theta_{01}$ conditioned on state $\ket{01}$ is $\theta_{01}=-\varphi_j$, giving $\theta_{10}-\theta_{00}<0$ for Bob.) However, Bob could use the same dispersive couplings as Alice to realize the entangling operation of Eq.~\eqref{app:entangling} (up to the same global phase of $\phi$), then redefine the states $\ket{s_2}\mapsto \ket{\bar{s}_2}$ of his register qubit (with $\bar{0}=1$ and $\bar{1}=0$) so that his entangling operation reads
\begin{equation}\label{app:entangling2}
    \ket{s_1s_2}\ket{\alpha}\mapsto\ket{s_1\bar{s}_2}\ket*{\alpha e^{i(\pi s_1+\varphi_js_2)}}.
\end{equation}
Equation \eqref{app:entangling2} gives $\theta_{00}=\varphi_j$, $\theta_{01}=0$, $\theta_{10}=\pi+\varphi_j$, and $\theta_{11}=\pi$, which realizes the required relation $\theta_{10}-\theta_{00}<0$ for Bob's entangling operation.

\section{Qubit dephasing}\label{sec:dephasing-appendix}

In an ideal scenario, a phase shift $\varphi_j$ is acquired by the light pulse conditioned on the state of Alice's $j^{\mathrm{th}}$ qubit being $\ket{1}$ [cf.~Eq.~\eqref{phase-breakdown}]. Under the assumption that $\varphi_j$ is set by tuning a dispersive coupling $\chi_j$, as described in Appendix \ref{sec:entangling-operation}, this scenario can be realized up to a global phase-space rotation by ensuring that [cf.~Eq.~\eqref{dispersive-shifts}]
\begin{equation}\label{phase-dispersive}
    \varphi_j=4\arctan{\left(\frac{2\chi_j}{\kappa_j}\right)}.
\end{equation}

For $\vert\Delta_j\vert>\kappa_j$, the dispersive coupling $\chi_j$ in Eq.~\eqref{phase-dispersive} is given by $\chi_j=g_j^2/\Delta_j$, where $g_j$ is the strength of the transverse coupling between qubit $j$ and its cavity (having frequency $\omega_{\mathrm{c}}$), and where $\Delta_j=\omega_{j}-\omega_{\mathrm{c}}$ is the qubit-cavity detuning. Quasistatic qubit dephasing is caused by shot-to-shot variation in the qubit frequency, $\omega_{j}\rightarrow \omega_{j}+\delta \omega_{j}$, which, in this instance, also leads to shot-to-shot variation of $\varphi_j$: $\varphi_j\rightarrow \varphi_j+\delta\varphi_j$. With $\varphi_j$ given by Eq.~\eqref{phase-dispersive}, an expression for $\delta\varphi_j$, valid to leading order in $\lvert\delta\omega_{j}/\Delta_j\rvert\ll 1$, is then given by
\begin{equation}\label{phase-dispersive-error}
    \delta\varphi_j=-2\sin{\left(\frac{\varphi_j}{2}\right)}\frac{\delta\omega_{j}}{\Delta_j}.
\end{equation}
The analogous result for Bob can be obtained by sending $\varphi_j\rightarrow -\varphi_j$.  

We first analyze the phase-qudit-mediated protocol of Sec.~\ref{sec:phase-qudit}, which involves a total of $2N$ stationary qubits. In a single shot, the state $\ket{\Psi_{\delta\omega}(t)}$ produced for a particular realization of qubit-frequency variations will be given by
\begin{equation}\label{state-dephasing}
    \ket{\Psi_{\delta\omega}(t)}=\frac{1}{2^N}\sum_{m,n}e^{i \Omega_{mn} t}\ket{m}_{\mathrm{A}}\ket{n}_{\mathrm{B}}\ket*{e^{i(m-n)\varphi}e^{i\Theta_{mn}}\alpha},
\end{equation}
where $\Omega_{mn}$ and $\Theta_{mn}$
are unwanted dynamical phases and phase shifts acquired by the qubits and light pulse, respectively, conditioned on the basis state $\ket{m}_{\mathrm{A}}\ket{n}_{\mathrm{B}}$ of Alice's and Bob's registers:
\begin{align}
\begin{aligned}\label{phimn}
    \Omega_{mn}&=\frac{1}{2}\sum_{j=0}^{N-1}\left[(-1)^{m_j}\delta\omega_{\mathrm{A},j}+(-1)^{n_j}\delta\omega_{\mathrm{B},j}\right]\\
    \Theta_{mn}&=-\sum_{j=0}^{N-1}\sin{\left(\frac{\varphi_j}{2}\right)}\left[(-1)^{m_j}\frac{\delta\omega_{\mathrm{A},j}}{\Delta_{\mathrm{A},j}}+(-1)^{n_j}\frac{\delta\omega_{\mathrm{B},j}}{\Delta_{\mathrm{B},j}}\right].
\end{aligned}
\end{align}
Here, $m_j=0,1$ labels the state of Alice's $j^{\mathrm{th}}$ qubit and is related to the decimal representation of the basis state $\ket{m}_{\mathrm{A}}$ through the relation $m=\sum_j 2^j m_j$. The variable $n_j=0,1$ similarly labels the state of Bob's $j^{\mathrm{th}}$ qubit and is related to $n$ through $n=\sum_j 2^j n_j$. We specify by the subscript `A' or `B' whether a qubit belongs to Alice or Bob.

Qubit-frequency variations consequently impact $\ket{\Psi_{\delta\omega}}$ in two ways: First, there are the usual dynamical phases (given by $\Omega_{mn}$) acquired by different computational basis states in the superposition. Averaging over these phases is what leads to quasistatic dephasing. However, qubit-frequency variations will also translate, under the model assumed in Appendix \ref{sec:entangling-operation}, to variations $\delta\varphi_j$ in the qubit-state-dependent phases $\varphi_j$ acquired by the light pulse as part of the entangling dynamics. Imperfections due to this latter effect will grow as the coherent-state amplitude $\alpha$ is increased.

As a measure of quality, we consider the average fidelity $\mathcal{F}_{\mathrm{phase}}(t)$ of $\ket{\Psi_{\delta\omega}}$ [Eq.~\eqref{state-dephasing}] relative to the ideal state $\ket{\Psi}$ [Eq.~\eqref{pre-meas-state}]:
\begin{equation}\label{definition-fidelity-dephasing}
\mathcal{F}_{\mathrm{phase}}(t)=\llangle \lvert \langle \Psi\vert \Psi_{\delta\omega}(t)\rangle\rvert^2\rrangle,   
\end{equation}
where here, $\llangle \rrangle$ denotes an average over the probability distribution governing the distribution of qubit-frequency variations. By evaluating the coherent-state overlaps, Eq.~\eqref{definition-fidelity-dephasing} can be rewritten as
\begin{equation}\label{fidelity-average}
    \mathcal{F}_{\mathrm{phase}}(t)=\frac{1}{16^N}\sum_{m,n,m',n'}\llangle e^{i\mathcal{Y}_{mnm'n'}(t)}\rrangle,
\end{equation}
where
\begin{equation*}
   \mathcal{Y}_{mnm'n'}=(\Omega_{mn}-\Omega_{m'n'})t+i\lvert\alpha\rvert^2\left(2-e^{i\Theta_{mn}}-e^{-i\Theta_{m'n'}}\right). 
\end{equation*}

For simplicity, we assume that $\delta\omega_{\mathrm{A},j}$ and $\delta\omega_{\mathrm{B},j}$ are Gaussian distributed with zero mean, and that variations across qubits are uncorrelated:
\begin{align}
\begin{aligned}\label{gaussian-dist}
    &\llangle \delta\omega_{\mathrm{A},j}\rrangle=\llangle \delta\omega_{\mathrm{B},j}\rrangle=0,\\
    &\llangle \delta\omega_{X,i}\delta\omega_{Y,j}\rrangle=\delta_{ij}\delta_{XY}\frac{2}{T_{2}^{*2}} 
\end{aligned}
\end{align}
Here, $\delta_{ij}$ is a Kronecker-delta function, and $T_2^*$ is a dephasing time here taken to be equal for all qubits. We expand $\mathcal{Y}_{mnm'n'}$ to quadratic order in $\Theta_{mn}$ (which is justifiable in the regime $\lvert\delta\omega_j/ \Delta_j\rvert\ll1$ considered here), and we perform a cumulant expansion of Eq.~\eqref{fidelity-average}, giving
\begin{equation}\label{cumulant-2}
    \mathcal{F}_{\mathrm{phase}}(t)=\frac{1}{16^N}\sum_{\substack{m,n\\m',n'}}e^{-\frac{\lvert\alpha\rvert^2}{2}\llangle \Theta_{mn}^2+\Theta_{m'n'}^2\rrangle-\frac{1}{2}\llangle \mathcal{Y}_{1,mnm'n'}^2(t)\rrangle},
\end{equation}
where
\begin{equation*}
 \mathcal{Y}_{1,mnm'n'}(t)=(\Omega_{mn}-\Omega_{m'n'})t+\lvert\alpha\rvert^2\left(\Theta_{mn}-\Theta_{m'n'}\right).
\end{equation*}
In order to evaluate Eq.~\eqref{cumulant-2}, we rewrite $\mathcal{Y}_{1,mnm'n'}$ as a sum over the qubit index $j$,
\begin{align}
\begin{aligned}
    \mathcal{Y}_{1,mnm'n'}(t)&=\frac{1}{2}\sum_{j=0}^{N-1} [(-1)^{m_j}-(-1)^{m_j'}]A_j(t)\delta\omega_{\mathrm{A},j}\\
    &+\frac{1}{2}\sum_{j=0}^{N-1}[(-1)^{n_j}-(-1)^{n_j'}]B_j(t)\delta\omega_{\mathrm{B},j},
\end{aligned}
\end{align}
where 
\begin{align}
    \mathcal{X}_j(t)=t-2\sin{\left(\frac{\varphi_j}{2}\right)} \frac{\lvert \alpha\rvert^2}{\Delta_{\mathcal{X},j}},\quad \mathcal{X}=\mathrm{A,B}.
\end{align}
Having expressed $\mathcal{Y}_{1,mnm'n'}(t)$ as a sum over uncorrelated variables, we can then use Eq.~\eqref{gaussian-dist} to straightforwardly evaluate its variance as
\begin{align}
\begin{aligned}\label{Y2}
    \llangle \mathcal{Y}_{1,mnm'n'}^2(t)\rrangle&=2\sum_{j}(1-\delta_{m_jm_j'})\frac{A_j^2(t)}{T_{2}^{*2}}+\begin{pmatrix}m\rightarrow n\\A\rightarrow B\end{pmatrix},\\
\end{aligned}
\end{align}
where $\delta_{m_jm_j'}$ is again a Kronecker-delta function. The short-hand appearing in the second term of Eq.~\eqref{Y2} represents a second, identical sum over $j$, up to the replacements $m\rightarrow n$ and $A\rightarrow B$. Since $\Theta_{mn}$ [Eq.~\eqref{phimn}] is already given as a sum over uncorrelated variables, we can also evaluate
\begin{align}\label{phi2}
    \Theta^2\equiv \llangle \Theta_{mn}^2\rrangle=\sum_{j=0}^{N-1}\sum_{\mathcal{X}=\mathrm{A,B}}  \sin^2\left(\frac{\varphi_j}{2}\right)\left(\Delta_{\mathcal{X},j}T_{2}^*\right)^{-2}.
\end{align}
Substituting Eqs.~\eqref{Y2} and \eqref{phi2} back into Eq.~\eqref{cumulant-2}, we can then group terms involving only Alice's or Bob's qubits and re-express $\mathcal{F}_{\mathrm{phase}}$ as a product of double sums over all basis states: 
\begin{equation}\label{fidelity-dephasing-2}
    \mathcal{F}_{\mathrm{phase}}(t)=\frac{e^{-\lvert\alpha\rvert^2\Theta^2}}{16^N}\prod_{\mathcal{X}=\mathrm{A,B}}\sum_{mm'}e^{-\sum\limits_{j}(1-\delta_{m_jm_j'})\frac{\mathcal{X}_j^2(t)}{T_{2}^{*2}}}.
\end{equation}

In a double sum of the form $\sum_{m=0}^{2^N-1}\sum_{m'=0}^{2^N-1}$, there are exactly $2^N$ terms where $m_j\neq m_j'$ for $k$ out of $N$ qubits for each value of $k=0,\dots,N$. (For instance, for each of the $2^N$ basis states $\ket{m}$ in the sum over $m$, there is exactly one basis state $\ket{m'=m}$ in the sum over $m'$ where $m_j=m_j'$ for all $j$, leading to $2^N$ terms in the double sum over $m$ and $m'$ where $m_j\neq m_j'$ for $k=0$ qubits.) We can therefore rewrite the double sum in Eq.~\eqref{fidelity-dephasing-2} as
\begin{equation}
    \sum_{mm'}e^{-\sum\limits_{j} (1-\delta_{m_jm_j'})(\dotsm)}=2^N\sum_{k=0}^N\sum_{\substack{\mathcal{S}\in \mathcal{S}_k}}e^{-\sum\limits_{j\in\mathcal{S}}(\dotsm)},\label{double-sum-sets}
\end{equation}
where $\sum_{\mathcal{S}\in\mathcal{S}_k}$ denotes a sum over sets $\mathcal{S}$ belonging to the set $\mathcal{S}_k$ of sets of cardinality $k$ defined as all $\binom{N}{k}$ possible combinations of $k$ integers drawn from $\{0,\dots, N-1\}$. As an example, for $N=3$ and $k=2$, the set $\mathcal{S}_{k=2}$ is given by all pairs of qubits: $\mathcal{S}_{2}=\{\{0,1\},\{0,2\},\{1,2\}\}$. 

In general, the inclusion of two ancilla qubits as part of the cat-qudit-mediated protocol will lead to additional dephasing.  A similar procedure can be used to calculate the average fidelity $\mathcal{F}_{\mathrm{cat}}(t)$ of the state $\ket{\Phi_{\delta\omega}(t)}$ produced in the presence of qubit-frequency variations relative to the ideal state $\ket{\Phi}$ given in Eq.~\eqref{final-cat-2}:
\begin{equation}
    \mathcal{F}_{\mathrm{cat}}(t)=\llangle\lvert\langle \Phi\vert \Phi_{\delta\omega}(t)\rangle\vert^2\rrangle.
\end{equation}
This quantity can be calculated by including ancilla-qubit-dependent dynamical phases and phase shifts in the cumulant expansion described previously. For concreteness, we assume that the ancilla qubits are coupled to their own cavities, in which case the analogue of Eq.~\eqref{phase-dispersive-error} is then  $\delta\varphi_{\mathcal{X},\mathrm{a}}=-2\delta\omega_{\mathcal{X},\mathrm{a}}/\Delta_{\mathcal{X},\mathrm{a}}$ for $\mathcal{X}=\mathrm{A,B}$ (Alice, Bob). In this scenario, dephasing of the $2N$ register qubits affects the state in the same manner as above, and it is straightforward to show that
\begin{equation}
    \mathcal{F}_{\mathrm{cat}}(t)=\mathcal{F}_{\mathrm{phase}}(t)\mathcal{G}_{\mathrm{a}}(t),
\end{equation}
where $\mathcal{G}_{\mathrm{a}}(t)$ quantifies additional imperfections due to ancilla-qubit dephasing with a functional form similar to that obtained for $\mathcal{F}_{\mathrm{phase}}$ [given in Eq.~\eqref{ga(t)} of the main text].

\section{Photon loss}\label{sec:full-loss}

In this section, we derive the action of the amplitude-damping channel (describing the effects of photon loss) on the light pulse transmitted from Alice to Bob, accounting for multiphoton losses occurring at higher order in the photon-loss probability. In the case of the cat-qudit-mediated protocol, we explicitly connect the action of the full loss model considered here to the simplified single-photon loss model of Sec.~\ref{sec:one-loss}, which we used to motivate the correction operator that should be applied to Alice's qubits conditioned on the parity eigenvalue $XX=-1$. We neglect any additional sources of photon loss apart from losses incurred by the light pulse while traveling from Alice to Bob.

\subsection{Phase-qudit-mediated protocol}

We begin by calculating the action of the amplitude-damping channel $\Lambda$ [Eq.~\eqref{amplitude-damping}] on the state $\rho_{\mathrm{A}}=\ketbra{\Psi}_{\mathrm{A}}$ [Eq.~\eqref{state-transmitted}] relevant to the phase-qudit protocol. Using the relation $\sqrt{\eta}^{\hat{a}^\dagger \hat{a}}\ket{\alpha}=e^{-(1-\eta)\frac{\lvert\alpha\rvert^2}{2}}\ket*{\sqrt{\eta}\alpha}$, we find that
\begin{equation}
    \Lambda(\rho_{\mathrm{A}})=\frac{1}{2^N}\sum_{m,m'} e^{\mathcal{Z}_{mm'}}\ketbra{m}{m'}\ketbra*{\sqrt{\eta}\alpha e^{im\varphi}}{\sqrt{\eta}\alpha e^{im'\varphi}},
\end{equation}
where, in terms of the average number $n_\ell=(1-\eta)\alpha^2$ of lost photons,
\begin{align}
\begin{aligned}
    &\mathcal{Z}_{mm'}(n_\ell)=i \phi_{mm'}(n_\ell)-\chi_{mm'}(n_\ell),\\
    &\phi_{mm'}(n_\ell)=n_\ell\sin{[(m-m')\varphi]},\\
    &\chi_{mm'}(n_\ell)=2n_\ell\sin^2{[(m-m')\frac{\varphi}{2}]}.
\end{aligned}
\end{align}
Loss-induced dephasing is therefore most severe for states with $m-m'=2^N/2$, corresponding to $(m-m')\varphi=\pi$. 

The light pulse encoding the phase qudit next interacts with Bob's qubits according to Eq.~\eqref{entangle-Bob}. Treating the measurement of the light pulse as an ideal projective measurement onto the basis $\{\ket{\sqrt{\eta}\alpha e^{-ik\varphi}}\}$, the post-measurement state of Alice's and Bob's qubits for a measurement of $\ket*{\sqrt{\eta}\alpha e^{-ik\varphi}}$ is then given by
\begin{equation}
    \rho_{\mathrm{phase},k}=\frac{1}{2^N}\sum_{m,m'}e^{\mathcal{Z}_{mm'}}\ketbra{m,m\oplus k}{m',m'\oplus k}.
\end{equation}
The fidelity $F_{\mathrm{phase}}=\langle \Psi_k\vert \rho_{\mathrm{phase},k}\vert \Psi_k\rangle$ of $\rho_{\mathrm{phase},k}$ relative to the ideal state $\ket{\Psi_k}$ [Eq.~\eqref{ideal-almost-bell}] can then be calculated as
\begin{equation}\label{fidelity-phase}
    F_{\mathrm{phase}}(n_\ell)=\frac{1}{4^N}\sum_{m,m'=0}^{2^N-1}e^{\mathcal{Z}_{mm'}(n_\ell)}.
\end{equation}
Since $\mathcal{Z}_{mm'}$ is a function only of the difference $m-m'$, we can simplify this expression by replacing the double sum over $m$ and $m'$ by a sum over a single variable $j$. This can be done by counting the number of ways a particular value of $\vert m-m'\rvert=j<2^N-1$ can be realized for $m,m'\in[0,2^N-1]$, leading to the identity
\begin{equation}\label{double-sum-replacement}
    \sum_{mm'}p(m-m')=2^N+\sum_{j=1}^{2^N-1}(2^N-j)\left[p(j)+p(-j)\right],
\end{equation}
valid when $p(m-m')$ is a function such that $p(0)=1$ and $p(m)=p(m+2^N)$. Here, the constant term $2^N$ comes from the fact that for every $m=0\dots, 2^N-1$, there is exactly one $m'$ in the double sum such that $m-m'=0$. The combinatorial factor $(2^N-j)$ in the sum over $j$ can be understood as reflecting the fact that for $m,m'\in[0,2^N-1]$, there are more combinations of $m$ and $m'$ such that $\vert m-m'\rvert=j$ for small $j$ compared to large $j$. By applying Eq.~\eqref{double-sum-replacement} to Eq.~\eqref{fidelity-phase}, we recover the expression given in Eq.~\eqref{fidelity-loss} of the main text for $\zeta=\mathrm{phase}$.

\subsection{Cat-qudit-mediated protocol}

We now calculate the action of the amplitude-damping channel $\Lambda$ [Eq.~\eqref{amplitude-damping}] on the state $\ket{\Phi}_{\mathrm{A}}$ of the light pulse and Alice's qubits [Eq.~\eqref{send-parity}] produced as part of the cat-qudit protocol. This is most straightforwardly done using the Fock-state representation of $M_k$ [Eq.~\eqref{krauss-operator}] together with the Fock-state representation of the even and odd cat states, 
\begin{align}
    \ket{C_{\alpha}^+}&=\frac{1}{\sqrt{\cosh{\lvert \alpha\rvert^2}}}\sum_{k=0}^\infty \frac{\alpha^{2k}}{\sqrt{(2k)!}}\ket{2k},\\
    \ket{C_{\alpha}^-}&=\frac{1}{\sqrt{\sinh{\lvert \alpha\rvert^2}}}\sum_{k=0}^\infty \frac{\alpha^{2k+1}}{\sqrt{(2k+1)!}}\ket{2k+1}.
\end{align}
Since $\ket{C_\alpha^\pm}$ are states of definite photon-number parity, it will be helpful to decompose $\Lambda$ into a sum of channels $\Lambda_\lambda$ resulting from an even ($\lambda=+$) or odd ($\lambda=-$) number of photon losses:
\begin{equation}
    \Lambda(\rho)=\Lambda_{+}(\rho)+\Lambda_{-}(\rho),
\end{equation}
where $\Lambda_{+,-}(\rho)=\sum_{k\:\mathrm{even,odd}}M_k \rho M_k^\dagger$. Since Alice's ancilla qubit will eventually be measured as part of an $XX$ parity check, we write $\Lambda_{\lambda}(\ketbra{\Phi}_{\mathrm{A}})$ as
\begin{equation}\label{channel-decomp}
    \Lambda_{\lambda}(\ketbra{\Phi}_{\mathrm{A}})=\sum_{\sigma,\sigma'=\pm}p_{\sigma\sigma'}^\lambda\tau_{\sigma\sigma'}^\lambda\ketbra{\sigma}{\sigma'},
\end{equation}
where here, $\ket{\sigma=\pm}$ is an $X$-basis eigenstate of Alice's ancilla and $\tau_{\sigma\sigma'}^\lambda$ is an operator acting on the light pulse and Alice's $N$ register qubits. In order to calculate $\tau_{\sigma\sigma'}^\lambda$, we first evaluate the action of the Kraus operators $M_k$ on the cat states $\ket{C_\alpha^\pm}$ for both even and odd $k$, giving 
\begin{align}
\begin{aligned}
    &M_{2k}\ket{C_\alpha^\pm}=\sqrt{\frac{J_\pm}{(2k)!}}\left(\sqrt{1-\eta}\alpha\right)^{2k}\ket*{C_{\sqrt{\eta}\alpha}^\pm},\\
    &M_{2k+1}\ket{C_\alpha^\pm}=\sqrt{\frac{K_\pm}{(2k+1)!}}\left(\sqrt{1-\eta}\alpha\right)^{2k+1}\ket*{C_{\sqrt{\eta}\alpha}^\mp},
\end{aligned}
\end{align}
where 
\begin{align}
\begin{aligned}
    J_+&=\frac{\cosh{(\eta\lvert\alpha\rvert^2)}}{\cosh{(\lvert \alpha\rvert^2)}},\\
    J_-&=\frac{\sinh{(\eta\lvert\alpha\rvert^2)}}{\sinh{(\lvert \alpha\rvert^2)}},\\
    K_+&=\frac{\sinh{(\eta\lvert\alpha\rvert^2)}}{\cosh{(\lvert \alpha\rvert^2)}},\\
    K_-&=\frac{\cosh{(\eta\lvert\alpha\rvert^2)}}{\sinh{(\lvert \alpha\rvert^2)}}.
\end{aligned}
\end{align} 
Summing over even and odd $k$ separately, we then obtain the coefficients $p_{\sigma\sigma'}^\pm$ appearing in Eq.~\eqref{channel-decomp},
\begin{align}
    \begin{aligned}
        p_{\sigma\sigma'}^{+}&=\frac{1}{4}\mathcal{N}_{ \alpha}^\sigma\mathcal{N_\alpha^{\sigma'}}\sqrt{J_\sigma J_{\sigma'}}\cosh{n_{\ell}},\\
        p_{\sigma\sigma'}^{-}&=\frac{1}{4}\mathcal{N}_{ \alpha}^\sigma\mathcal{N_\alpha^{\sigma'}}\sqrt{K_\sigma K_{\sigma'}}\sinh{n_{\ell}},\\
    \end{aligned}
\end{align}
from which it may be verified that $\sum_{\sigma}\sum_\lambda p_{\sigma\sigma}^{\lambda}=1$ as required for $\mathrm{Tr}\:\Lambda(\ketbra{\Phi}_{\mathrm{A}})=1$. As before, $\mathcal{N}_\alpha^\pm=[2(1\pm e^{-2\lvert\alpha\rvert^2})]^{1/2}$ are normalization factors appearing in the definition $\ket{C_\alpha^\pm}=(\mathcal{N}_\alpha^\pm)^{-1}(\ket{\alpha}\pm\ket{-\alpha})$.  This re-summation also gives
\begin{align}\label{states-even-odd}
\begin{aligned}
    \tau_{\sigma\sigma'}^{\pm} &= \frac{1}{2^N}\sum_{m,m'}\Omega_{mm'}^{\pm}\ketbra*{m,c_m^{\pm\sigma}}{m',c_{m'}^{\pm\sigma'}},\\
\end{aligned}
\end{align}
where for compactness, we have introduced the notation $\ket{m,c_m^\sigma}\equiv \ket{m}_{\mathrm{A}}\ket*{C_{\sqrt{\eta}\alpha e^{im\varphi}}^\sigma}$, as well as a parameter $\Omega_{mm'}^{\lambda}\in [0,1]$ quantifying the amount of dephasing between basis states $\ket{m}$ and $\ket{m'}$ resulting from the action of the loss channel $\Lambda_\lambda$:
\begin{align}
    &\Omega_{mm'}^{+}=\frac{\cosh{(n_{\ell} e^{i(m-m')\varphi})}}{\cosh{n_{\ell}}},\\
    &\Omega_{mm'}^{-}=\frac{\sinh{(n_{\ell}e^{i(m-m')\varphi})}}{\sinh{n_{\ell}}}.\label{odd-dephasing-suppression}
\end{align}
The expression for $\Omega_{mm'}^{-}$ can be connected back to the simplified single-photon loss model leading to Eq.~\eqref{simplified-loss} by observing that to leading order in $n_{\ell}$, $\Omega_{mm'}^{-}=e^{i(m-m')\varphi}+O(n_{\ell}^2)$. This is the same $m$-dependent phase factor as that appearing in Eq.~\eqref{simplified-loss}. 



Following the interaction of the light pulse with Bob's qubits [cf.~Eq.~\eqref{entangle-bob-2}], the state $\rho$ of the light pulse and all stationary qubits (including the two ancillas) can be written as
\begin{equation}\label{mixed-parity}
    \rho=\sum_{\lambda=\pm} p^\lambda \rho^\lambda,
\end{equation}
where $p^\lambda=\sum_{\sigma=\pm}p_{\sigma\sigma}^\lambda$. In Eq.~\eqref{mixed-parity}, the states $\rho^{\lambda=\pm}$ are states of definite $XX=\lambda$ parity (for the ancilla qubits) given by
\begin{align}
\begin{aligned}
    &\rho^\pm=\frac{1}{p^\pm}\sum_{\sigma,\sigma'=\pm}p_{\sigma\sigma'}^{\pm}\rho_{\sigma\sigma'}^{\pm}\ketbra{\sigma(\pm\sigma)}{\sigma'(\pm\sigma')}.\\
\end{aligned}
\end{align}
Here, $\ket{\sigma(\pm\sigma)}=\ket{\sigma}\otimes \ket{\pm\sigma}$ is a state of the ancilla qubits, while $\rho_{\sigma\sigma'}^\pm$ is an operator acting on the light pulse and $2N$ register qubits held by Alice and Bob, given by
\begin{align*}
\begin{aligned}
    \rho_{\sigma\sigma'}^\pm&=\frac{1}{4^N}\sum_{\substack{m,m'\\n,n'}}\Omega_{mm'}^{\pm}\ketbra*{m,n,c_{m-n}^{\pm\sigma}}{m',n',c_{m'-n'}^{\pm\sigma'}},\\
\end{aligned}
\end{align*}
where here, $\ket{m,n,c_{k}^\sigma}\equiv\ket{m}_{\mathrm{A}}\ket{n}_{\mathrm{B}}\ket*{C_{\sqrt{\eta}\alpha e^{ik\varphi}}^\sigma}$.
Given $\rho$ [Eq.~\eqref{mixed-parity}], an $XX$ parity check of the two ancilla qubits will return a state of definite $XX=\pm 1$ parity, given by either $\rho^+$ or $\rho^-$. Once $\rho^\pm$ has been obtained, independent $Z$-basis measurements of the two ancilla qubits yielding the measurement outcomes $ss'=00,01,10,11$ will produce the state
\begin{equation}\label{post-meas-z}
    \rho^{\lambda,\mu}=\frac{\Tilde{\rho}^{\lambda,\mu}}{\mathrm{Tr}[\Tilde{\rho}^{\lambda,\mu}]}, \quad \mu=s+s'\;(\mathrm{mod}\;2),
\end{equation}
where

\begin{equation}
    \Tilde{\rho}^{\lambda,\mu}=\sum\limits_{\sigma=\pm}p_{\sigma\sigma}^{\lambda}\rho_{\sigma\sigma}^{\lambda}+(-1)^{\mu}\sum\limits_{\sigma=\pm}p_{\sigma(-\sigma)}^{\lambda}\rho_{\sigma(-\sigma)}^{\lambda}.
\end{equation}
For each of the two possible parity eigenvalues $XX=\lambda=\pm 1$, the post-measurement state $\rho^{\lambda,\mu}$ [Eq.~\eqref{post-meas-z}] produced by the $Z$-basis measurements depends only on the parity $\mu=0,1$ of the measurement outcomes $s$ and $s'$.

We model the heterodyne detection of the light pulse as a projection onto $\{\ket*{\sqrt{\eta}\alpha e^{-ik\varphi}}\}$, effectively treating the overlap between different coherent states as negligible for $k\neq k'$. Given $\rho^{\lambda,\mu}$ and the heterodyne measurement outcome $\ket{\sqrt{\eta}\alpha e^{-ik\varphi}}$, the post-measurement state $\rho_k^{\lambda,\mu}$ of Alice's and Bob's $2N$ qubits can then be written as
\begin{equation}\label{obtained-state}
    \rho_k^{\lambda,\mu}=\frac{1}{2^N}\sum_{m,m'}\Omega_{mm'}^\lambda\ketbra*{\Psi_{m,k}^{\lambda,\mu}}{\Psi_{m',k}^{\lambda,\mu}},
\end{equation}
where
\begin{align}
\begin{aligned}\label{uglystates}
    \ket*{\Psi_{m,k}^{\pm,\mu}}&=D^{(-)^\mu}\ket{m,m\oplus k}\\
    &\pm D^{(-)^{\mu+1}}\ket{m,m\oplus(k+2^{N-1})}.\\
\end{aligned}
\end{align}
Here, $D^\pm=\frac{1}{2}\left(\sqrt{1+e^{-2\eta \alpha^2}}\pm \sqrt{1-e^{-2\eta\alpha^2}}\right)$ are coefficients satisfying $(D^+)^2+(D^-)^2=1$.

Suppose now that the ancilla-qubit measurements returned a parity-check eigenvalue of $XX=\pm 1$ and an \textit{even} parity for the sum of the independent $Z$-basis measurement outcomes, the latter corresponding to the case $\mu=0$. In this scenario, the ideal state obtained in the absence of errors would be $\ket{\Psi_k}$ for $XX=+1$ and $R_{\mathrm{A}}^\dagger\ket{\Psi_k}$ for $XX=-1$ [cf.~Eq.~\eqref{simplified-loss}]. For $\mu=1$, it would instead be $\ket{\Psi_{k+2^{N-1}}}$ for $XX=+1$ and $R_{\mathrm{A}}^\dagger\ket{\Psi_{k+2^{N-1}}}$ for $XX=-1$, and the following calculation would proceed similarly. To quantify the errors due to multiphoton losses, we calculate the fidelity $F_{\mathrm{cat}}^+=\langle \Psi_k\vert \rho_k^{+,0}\vert \Psi_k\rangle$ of the state $\rho_k^{+,0}$ [Eq.~\eqref{obtained-state}] obtained upon measuring $XX=+1$ and $\mu=0$, relative to the ideal state
$\ket{\Psi_k}$ [Eq.~\eqref{ideal-almost-bell}], as well as the fidelity $F_{\mathrm{cat}}^-=\langle \Psi_k\vert R_{\mathrm{A}} \rho_k^{-,0} R_{\mathrm{A}}^\dagger\vert \Psi_k\rangle$ of the state $\rho_k^{-,0}$ obtained upon measuring $XX=-1$ and $\mu=0$, relative to the ideal state $R_{\mathrm{A}}^\dagger\ket{\Psi_{k}}$. Using Eqs.~\eqref{obtained-state} and \eqref{uglystates}, these can be written as
\begin{align}
    F_{\mathrm{cat}}^+(n_\ell) &=\frac{1}{4^N}(D^+)^2\sum_{m,m'}\Omega_{mm'}^{+}(n_\ell),\label{cat-fid-even}\\
    F_{\mathrm{cat}}^-(n_\ell)&=\frac{1}{4^N}(D^+)^2\sum_{m,m'}e^{-i(m-m')\varphi}\Omega_{mm'}^{-}(n_\ell),\label{cat-fid-odd}
\end{align}
where $(D^+)^2=(1+\sqrt{1-e^{-4\eta\alpha^2}})/2$ is independent of $n_{\ell}$ and can reasonably be set to 1 provided the average number $\eta\alpha^2$ of photons transmitted is $\eta\alpha^2\gg 1$. Applying Eq.~\eqref{double-sum-replacement} to Eqs.~\eqref{cat-fid-even} and \eqref{cat-fid-odd} then allows us to write
\begin{equation}\label{two-cats}
    F_{\mathrm{cat}}^\pm(n_\ell) =\frac{1}{4^N}\left[2^N+\sum_{j=1}^{2^N-1}(2^N-j)f_{\mathrm{cat}}^\pm(n_\ell,j)\right],
\end{equation}
where
\begin{align*}
\begin{aligned}
    f_{\mathrm{cat}}^+(n_\ell,j)&=2\cos{(n_\ell \sin{j\varphi})}\frac{\cosh{(n_\ell\cos{j\varphi})}}{\cosh{n_\ell}},\\
    f_{\mathrm{cat}}^-(n_\ell,j)&=\sum_{\sigma=\pm 1}\mathrm{sgn}(\sigma)\frac{e^{\sigma n_\ell \cos{j\varphi}}\cos{(n_\ell\sin{j\varphi}-\sigma j\varphi)}}{\sinh{n_\ell}}.
\end{aligned}
\end{align*}
By expanding $f_{\mathrm{cat}}^\pm(j)$ in powers of $n_\ell$ and performing the sum over $j$ in Eq.~\eqref{two-cats}, we then find that, independent of $N$, 
\begin{align}
    F_{\mathrm{cat}}^+(n_\ell)&=1-\frac{n_\ell^2}{2}+O(n_\ell^4),\label{eq:ConditionalFPlus}\\
    F_{\mathrm{cat}}^-(n_\ell)&=1-\frac{n_\ell^2}{6}+O(n_\ell^4).\label{eq:ConditionalFMinus}
\end{align}

The conditional fidelities $F_{\mathrm{cat}}^\pm$ given in Eqs.~\eqref{eq:ConditionalFPlus} and \eqref{eq:ConditionalFMinus} can also be understood by considering the binomial distribution for the probability $P(n')$ of losing $n'$ photons given a transmission probability $\eta$ per photon and a typical number of photons $n$ passing from Alice to Bob:
\begin{equation}\label{eq:BinomialDistribution}
P(n') = \binom{n}{n'}\eta^{n-n'}(1-\eta)^{n'}.
\end{equation}
A syndrome measurement $XX=+1\,(XX=-1)$ signals that $n'$ is even (odd). If we assume the protocol is perfect (error free) for $n'=0$ or $n'=1$ lost photons (after applying the appropriate correction operation), but leads to a state with zero overlap with the intended target for $n'\ge 2$, then the conditional infidelities can be written as
\begin{eqnarray}
1-F_{\mathrm{cat}}^+ & = & \frac{\sum_{n'=2,4,\ldots}P(n')}{\sum_{n'=0,2,4,\ldots}P(n')},\\
1-F_{\mathrm{cat}}^- & = & \frac{\sum_{n'=3,5,\ldots}P(n')}{\sum_{n'=1,3,5,\ldots}P(n')}.
\end{eqnarray}
In the limit of a small per-photon loss probability, $1-\eta \ll 1$, we have $P(n'+2)\ll P(n')$, yielding the approximate infidelities 
\begin{eqnarray}
1-F_{\mathrm{cat}}^+ & \simeq & P(2)/P(0),\label{eq:FcatPlusApprox}\\
1-F_{\mathrm{cat}}^- & \simeq & P(3)/P(1).\label{eq:FcatMinusApprox}
\end{eqnarray}
If we additionally assume $n\gg 1$ and $n_\ell=n(1-\eta)\ll 1$, then the binomial probability distribution $P(n')$ can be approximated by a Poisson distribution:
\begin{equation}\label{eq:FishyDistribution}
P(n') \simeq \frac{n_\ell^{n'} e^{-n_\ell}}{n'!}.
\end{equation}
Inserting the Poissonian approximation for $P(n')$ into Eqs.~\eqref{eq:FcatPlusApprox} and \eqref{eq:FcatMinusApprox} then directly recovers the results from Eqs.~\eqref{eq:ConditionalFPlus} and \eqref{eq:ConditionalFMinus}:
\begin{eqnarray}
1-F_{\mathrm{cat}}^+(n_\ell) & \simeq & n_\ell^2/2,\\
1-F_{\mathrm{cat}}^-(n_\ell) & \simeq & n_\ell^2/6.
\end{eqnarray}
If we assume perfect correction operations, the conditional infidelity due to photon loss is actually three times \emph{lower} when there is a heralded photon-loss event ($XX=-1$) compared to when there is no heralded loss ($XX=+1$). This can be traced back to the rapid (super-exponential) decay of the Poisson distribution, Eq.~\eqref{eq:FishyDistribution}, with increasing $n'$. This behavior leads to a smaller relative probability to lose three photons (relative to the probability to lose one) compared to the relative probability to lose two photons (relative to the probability to lose zero). 

\bibliography{bib}

\end{document}